\documentclass[apj]{emulateapj}

\usepackage{subfigure}
\usepackage[dvipsnames]{xcolor}

\newcommand{\myemail}{s.w.jones@keele.ac.uk}

\slugcomment{Submitted to ApJ 2013 February 19; accepted 2013 June 4}

\shorttitle{Advanced burning stages and fate $8-10\,M_\odot$ stars}
\shortauthors{Jones et al.}

\begin{document}

\title{Advanced burning stages and fate of $8-10\,M_\odot$ stars}

\author{S. Jones\altaffilmark{1,*}, R. Hirschi\altaffilmark{1,2}, K.
Nomoto\altaffilmark{2}, T. Fischer\altaffilmark{3,4}, F.X. Timmes\altaffilmark{5,6},
F. Herwig\altaffilmark{7,6}, B. Paxton\altaffilmark{8}, H. Toki\altaffilmark{9}, T. Suzuki\altaffilmark{10,11}, G. Mart\'{i}nez-Pinedo$^{4,3}$, Y. H. Lam$^{4}$,  M. G. Bertolli\altaffilmark{12}}
\affil{$^1$ Astrophysics Group, Lennard Jones Building, Keele University ST5
5BG, UK \\
$^2$ Kavli Institute for the Physics and Mathematics of the Universe (WPI), The University of Tokyo, Kashiwa, Chiba 277-8583, Japan \\
$^3$ GSI Helmholtzzentrum f{\"u}r Schwerionenforschung GmbH, Planckstra§e 1, 64291, Darmstadt, Germany \\
$^4$ Institut f{\"u}r Kernphysik, Technische Universit{\"a}t Darmstadt, Schlossgartenstra{\ss}e 2, 64289 Darmstadt, Germany. \\
$^5$ School of Earth and Space Exploration, University of Arizona, Tempe, AZ
85287, USA \\
$^6$ Joint Institute for Nuclear Astrophysics, University of Notre Dame, IN
46556, USA \\
$^7$ Department of Physics and Astronomy, Victoria, BC V8W 3P6, Canada \\
$^8$ KITP and Dept. of Physics, University of California, Santa Barbara, CA
93106 USA \\
$^9$ Research Center for Nuclear Physics (RCNP), Osaka University, Osaka 567-0047, Japan \\
$^{10}$ Department of Physics, College of Humanities and Sciences, Nihon University
Sakurajosui 3-25-40, Setagaya-ku, Tokyo 156-8550, Japan \\
$^{11}$ National Astronomical Observatory of Japan, Mitaka, Tokyo 181-8588, Japan \\
$^{12}$ Theoretical Division, Los Alamos National Laboratory, Los Alamos, NM 87545, USA \\
}

\altaffiltext{*}{email: \myemail}

\begin{abstract}

The stellar mass range $8\lesssim M/M_\odot \lesssim12$ corresponds to
the most massive AGB stars and the most numerous massive stars. It is host to a variety of supernova progenitors and is therefore very important for galactic chemical evolution and stellar population studies.
In this paper, we study the transition from super-AGB star to massive star and find that a propagating neon-oxygen burning shell is common to both the most massive electron capture supernova (EC-SN) progenitors and the lowest mass iron-core collapse supernova (FeCCSN) progenitors.
Of the models that ignite neon burning off-center, the $9.5\,M_\odot$ star would evolve to an FeCCSN after the neon-burning shell propagates to the center, as in previous studies. The neon-burning shell in the $8.8\,M_\odot$ model, however, fails to reach the center as the URCA process and an extended ($0.6\,M_\odot$) region of low $Y_\mathrm{e}$ (0.48) in the outer part of the core begin to dominate the late evolution; the model evolves to an EC-SN. This is the first study to follow the most massive EC-SN progenitors to collapse, representing an evolutionary path to EC-SN in addition to that from SAGB stars undergoing thermal pulses. We also present models of an $8.75\,M_\odot$ super-AGB star through its entire thermal pulse phase until electron captures on $^{20}\mathrm{Ne}$ begin at its center and of a $12\,M_\odot$ star up to the iron core collapse. 
We discuss key uncertainties and how the different pathways to collapse affect the pre-supernova structure. Finally, we compare our results to the observed neutron star mass distribution.
\end{abstract}


\section{Introduction}
As helium fuel is exhausted at their center, stars with initial masses
$M\gtrsim1\,M_\odot$ develop cores consisting of mostly carbon and oxygen (CO).
These CO cores become partially degenerate in stars with
$M\lesssim9\,M_\odot$ before the threshold temperature for carbon ignition can be
reached at the center. Neutrino processes cause a temperature inversion in the
core and if the star is massive enough ($6\lesssim M/M_\odot \lesssim 9$),
carbon ignites off-center and proceeds to burn inward (see
e.g. \citealp{Nomoto1984,Garcia-Berro1997,Siess2007}).
After the core has been processed by carbon burning, it consists of mostly
oxygen and neon (ONe) in a degenerate configuration. Stars experiencing the
off-center ignition of carbon to form a degenerate ONe core are known as
super-AGB (SAGB) stars.

Efforts to better understand the evolution of SAGB stars through numerical
modeling are ongoing \citep[see e.g.][]{Siess2010,Doherty2010} and it is now
computationally possible to follow several thousands of thermal pulses in order
to explore the complex evolution that can be compared with observations. The shortcomings of hydrostatic 1-D modelling of this phase were
recently briefly discussed by \citet{Lau2012}.

$8-12\,M_\odot$ stars are of crucial importance to galactic chemical
evolution and stellar population studies. SAGB stars are at the lower
end of this mass range, whose massive envelopes
enrich the interstellar medium.
At the upper
end of this mass range are the most abundant of the massive stars (because the IMF is bottom-heavy). These massive stars produce violent explosions in their deaths, producing and expelling heavier elements.
The statistical contribution of stars in this mass range to supernovae and their remnants is well reflected in the derived progenitor mass distribution of M31 \citep{Jennings2012}. The authors found the IMF of M31 to be steeper than the $\alpha=-2.35$ of Salpeter when assuming a single power law.

Super-AGB stars also
have the potential to explode as electron-capture supernovae (EC-SNe) if their ONe core masses grow large enough to develop central densities in excess of the threshold density for $^{20}\mathrm{Ne}(\mathrm{e}^-,\nu)^{20}\mathrm{F}$
\citep{Miyaji1980,Nomoto1984,Nomoto1987,Miyaji1987,Gutierrez1996,Ritossa1999,Poelarends2008}.
Such an explosion is thought to have given birth to the Crab Nebula
\citep{Davidson1982,Nomoto1982crabnature,Wanajo2009}. EC-SNe have also been
proposed as potential sites for the \emph{r}-process \citep{Ning2007},
responsible for the synthesis of the heaviest elements. However more recent studies
of the shocked surface layers in the explosion do not produce the conditions
necessary to create such heavy elements in great abundance \citep{Hoffman2008,Wanajo2009}. Instead, the explosive nucleosynthesis calculations so far only predict a contribution to the lighter nuclei, Zn and Zr for example.
Increasing evidence suggests that there may be weak and main components to the $r$-process \citep{Hansen2012}, of which EC-SN events could contribute the weak component.

Stars in this mass range may hold the key to explaining observations of
sub-luminous type II-P
supernovae with low $^{56}\mathrm{Ni}$ ejecta \citep{Smartt2009}. Interestingly, there is a non-monotonic feature of the progenitor star mass--luminosity relation in this mass range, where the occurrence of deep second dredge-up (2DUP) boosts the luminosity of super-AGB stars. The luminosity of super-AGB stars can then come close to exceeding even that of a $15\,M_\odot$ star \citep[see e.g.][]{Eldridge2007}.

\citet[][case 2.2]{Nomoto1987} and \citeyearpar[][case 2.4]{Nomoto1984} has provided the
canonical pre-SN structures for EC-SN simulations (\citealp{Kitaura2006,Fischer2010}
and for the resulting nucleosynthesis see \citealp{Wanajo2009,Wanajo2011}).
The model from \citet{Nomoto1984} dubbed case
2.6 was not followed any further than the ignition of off-center
neon and oxygen burning shells. Subsequently, \citet{NomotoHashimoto1988} followed
the propagation of the neon-burning shell in a helium star of $3.0\,M_\odot$ (case 3.0) to the stellar center, concluding that the star
would produce an Fe core before collapsing.
\citet{Timmes1992} and \citet{Timmes1994} studied in detail
the properties of nuclear flames in degenerate compositions of C+O and O+Ne+Mg.
In these studies it was proposed that, should neon and oxygen burning ignite
off-center in the
core of a star significantly far from the center, then it may compete with the
contraction of the center to determine its fate --- EC-SN or FeCCSN.
Currently there are no progenitor models for this additional path to EC-SN.
We call these failed massive stars and present them for the first time in this paper.
The subtle differences between the pre-SN
evolution of these progenitors and SAGB progenitors could affect
the explosion \citep{Gutierrez1996}.
More recently, \citet{EldridgeTout2004} reported that for the most massive
SAGB stars, an ONe core with $M>M_\mathrm{Ch}$ is produced before the completion of the
second dredge-up, and subsequently its mass is reduced to $M_\mathrm{Ch}$ by the ignition of a carbon-burning convective shell. It is unclear from this study what then happens to these stellar models as the evolution was not calculated any further and a limited nuclear reaction network was used for the calculation.

The goal of this paper is thus to present evolutionary models and progenitor structures for both SAGB and failed massive stars and discuss the possible impact of this additional channel on EC-SNe.
The structure of this paper is as follows. In \S\,\ref{methodsec}, the input physics for the models is described; \S\,\ref{fateevsec} contains detailed description of the evolution of our models and their fate.
The neon-oxygen shell burning in the 8.8 and $9.5\,M_\odot$ (\S\,\ref{lateev} and \ref{lateev2}) and the progenitor structures of our models (\S\,\ref{progenitorstructures}) are also presented. \S\,\ref{discussion} discusses the key uncertainties in modelling stars in the $8-12\,M_\odot$ mass range and finally, we summarize our results and conclude in \S\,\ref{conclusions}.

\section{Methodology and models}
\label{methodsec}
We calculated stellar models with initial masses of $8.2$, $8.7$, $8.75$, $8.8$, $9.5$ and $12.0\,M_\odot$ with the \emph{Modules for Experiments in Stellar Astrophysics} (MESA) stellar evolution code \citep{MESA2011}, revision 3709. We calculated non-rotating models from the pre-main sequence assuming a uniform initial composition with a metal fraction of $Z=0.014$ and
elemental abundances taken from \citet{AGS2004}.

In MESA, convective mixing is treated as a
time-dependent, diffusive process with a diffusion
coefficient, $D_{\rm MLT}$.  See \citet{MESA2011} for 
the implementation details of standard mixing length treatment.
The mixing length parameter is chosen as $\alpha_\mathrm{MLT}=1.73$ from fitting of the parameters of the Sun.
During the entire evolution sequence we assume the Schwarzschild
criterion for convection with the exception of the late stages of the $8.75\,M_\odot$ and $8.8\,M_\odot$ models (when electron
captures begin to dominate the evolution of the core) where we employ the Ledoux criterion
\citep{Miyaji1987}.
Mixing at convective boundaries is treated with an
exponentially decaying diffusion coefficient \citep{Freytag1996,Herwig2000} of
the form
\begin{equation}
 D=D_0\mathrm{exp}\bigg{(}-\frac{2z}{f_\mathrm{CBM}\lambda_{P,0}}\bigg{)}
 \label{exponentialovershoot}
\end{equation}
where
$D_0$ is the diffusion coefficient, taken equal to the mixing length diffusion coefficient value ($D_\mathrm{MLT}$) at a distance $f_\mathrm{CBM}\lambda_\mathrm{P,S}$ inside the convection zone from the Schwarzschild boundary. At this location, the pressure scale height is $\lambda_\mathrm{P,0}$, while $\lambda_\mathrm{P,S}$ is the pressure scale height at the Schwarzschild boundary. This is because the value of $D_0$ drops sharply towards zero at the Schwarzschild boundary.
$D$ is the diffusion coefficient as a function of distance $z$ from this location and $f_\mathrm{CBM}$ is a free parameter, for which we assume the value of
0.014 at all convective boundaries except for at the base of convective shells
burning nuclear fuel, for which we assume a stricter value of
$f_\mathrm{CBM}=0.005$. Such a reduced efficiency of convective boundary mixing at the bottom of shell-flash convection zones is indicated from both He-shell flash convection in AGB stars \citep{HerwigAGB2005} as well as nova shell flashes \citep{Denissenkov2013}. During the silicon-burning stage of the $12\,M_\odot$ model, no convective boundary mixing is assumed ($f_\mathrm{CBM}=0$). Future 3D simulations are required to constrain the behavior of convective boundary mixing during these late stages.

MESA solves the coupled stellar structure, nuclear burning,
and abundance mixing equations simultaneously.
In cases where the burning timescale is much longer than the mixing
timescale, as for example during core H-burning on the main sequence, then MESA's coupled calculation and an operator-split
calculation will agree.  
In cases where the nuclear burning time scale is similar or shorter
compared to the mixing time scale, the coupled method provides consistent abundance profiles in convection
zones, whereas operator-split calculations require a special treatment for chemical species with short nuclear timescales and smaller time steps. Note that in exceptional cases where the energy 
release by simultaneous burning and mixing is so large that 
the approximations of MLT are violated, then all 1D methods become inaccurate and 3D hydrodynamic simulations are necessary \citep[e.g.][]{Herwig2011}.

\begin{figure*}
  \centering
  \includegraphics[width=0.9\textwidth]{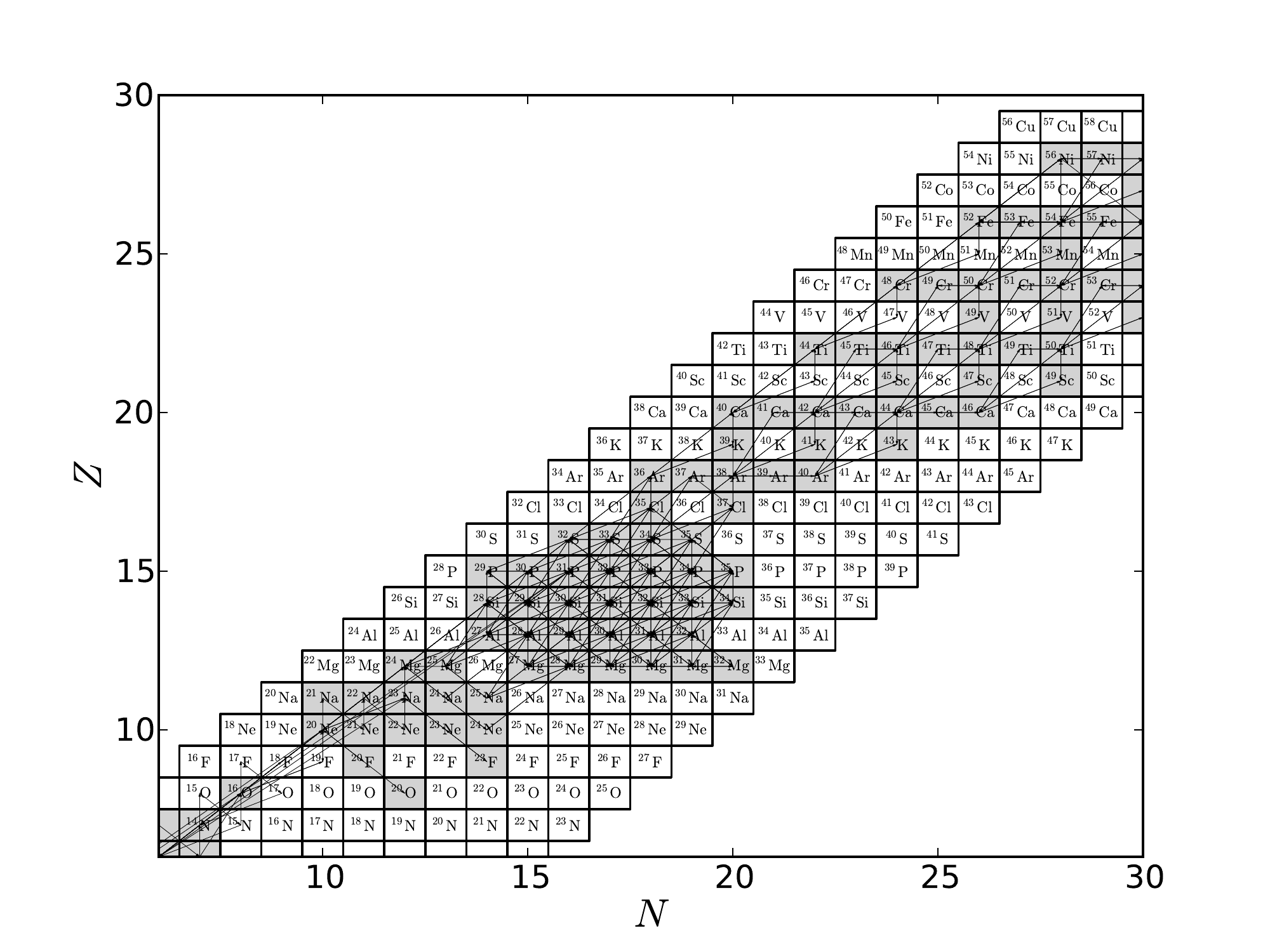}
  \caption{The nuclear reaction network used in these calculations. The abundances of the shaded isotopes are followed explicitly. Reactions are shown with arrows, and implicitly included isotopes (ones whose abundances we do not calculate explicitly but through which reactions are considered to proceed) can be seen where there is an arrow junction on an unshaded isotope. This network was used for all of the models with the exception of the TP-SAGB phase in the 8.2, 8.7 and 8.75\,M$_\odot$ models and the post oxygen-burning phase in the 12\,M$_\odot$ model. In these phases, appropriate smaller, more efficient networks are used.}
  \label{network_plot}
\end{figure*}

We trace the nuclear energy production and composition evolution with a network
of 114 isotopes from $^1\mathrm{H}$ to $^{61}\mathrm{Co}$ including the NeNa cycle,
URCA processes, alpha chains and electron-captures by $^{24}\mathrm{Mg}$,
$^{24}\mathrm{Na}$, $^{20}\mathrm{Ne}$ and $^{20}\mathrm{F}$ along with their
inverses. Fig.\,\ref{network_plot} shows the detail of the network. Such a large network is required to follow both nucleosynthesis and
energy production in these models. For example $^{30}\mathrm{Si}$ and $^{34}\mathrm{S}$
are the main products of O-burning in the lowest--mass massive stars as opposed to
$^{28}\mathrm{Si}$ and $^{32}\mathrm{S}$ in more massive stars owing to higher
degeneracies and thus higher electron capture rates \citep[see e.g.][]{ThielemannArnett1985}. In stars with degenerate cores close to the Chandrasekhar limit, accurately calculating the electron fraction, $Y_\mathrm{e}$, is very important because only a slight reduction in $Y_\mathrm{e}$ can cause significant contraction.
Further isotopes are included implicitly to account for non-negligible reaction channels, for example $^{44}\mathrm{Ti}(\alpha,p)^{47}\mathrm{V}(p,\gamma)^{48}\mathrm{Cr}$ is included though we do not explicitly calculate the abundance of $^{47}\mathrm{V}$. These implicit isotopes can be seen in Fig.\,\ref{network_plot} where there is an arrow junction on an unshaded isotope.
For the 8.2, 8.7 and $8.75\,M_\odot$ models that become SAGB stars, a network optimised for the AGB phase, including 37 isotopes and the relevant nuclear processes listed above, was employed from the time of completion of second dredge-up.
During the silicon-burning stage of the $12\,M_\odot$ model, we employ the simplified 21-isotope network, approx21.net, that is available in the MESA code. It is common for simplifications to the nuclear reaction network to be made in order to efficiently deal with the many high rates of forward and reverse reactions.
Weak reaction rates and associated neutrino--loss rates are those of
\citet{FFNweak1985}, \citet{Takahara89}, \citet{ODA94}, \citet{Langanke2000} and, as will be discussed in \S\S\,\ref{massivestarsection} and \ref{weakdiscussion}, \citet{Toki2013}.
Assumed mass--loss rates comprise that of
\citet[][$\eta=0.5$]{Reimers1975} for the red giant branch (RGB) phase and
\citet[][$\eta=0.05$]{Bloecker1995} during the AGB phase.

\section{Evolution and fates}
\label{fateevsec}
In this section, the evolution and fate of the models are described in the following order. In \S\,\ref{earlyev} the early evolution of the models from the main sequence to the end of carbon burning is briefly outlined. \S\S\,\ref{SAGBsection} and \ref{massivestarsection} then describe in detail the late evolution of the super-AGB and massive star models, respectively, wherein \S\S\,\ref{lateev} and \ref{lateev2} describe the behavior of neon-oxygen burning shells in the 8.8 and $9.5\,M_\odot$ models. Lastly, in \S\,\ref{progenitorstructures} the progenitor structures of our models are described, comparing both between models calculated for this study and with other progenitor models currently published in this mass range \citep{Nomoto1987,Woosley2002zz}.

\subsection{Evolution to the end of carbon burning}
\label{earlyev}
\begin{figure*}
  \centering
  \includegraphics[height=0.425\textheight]{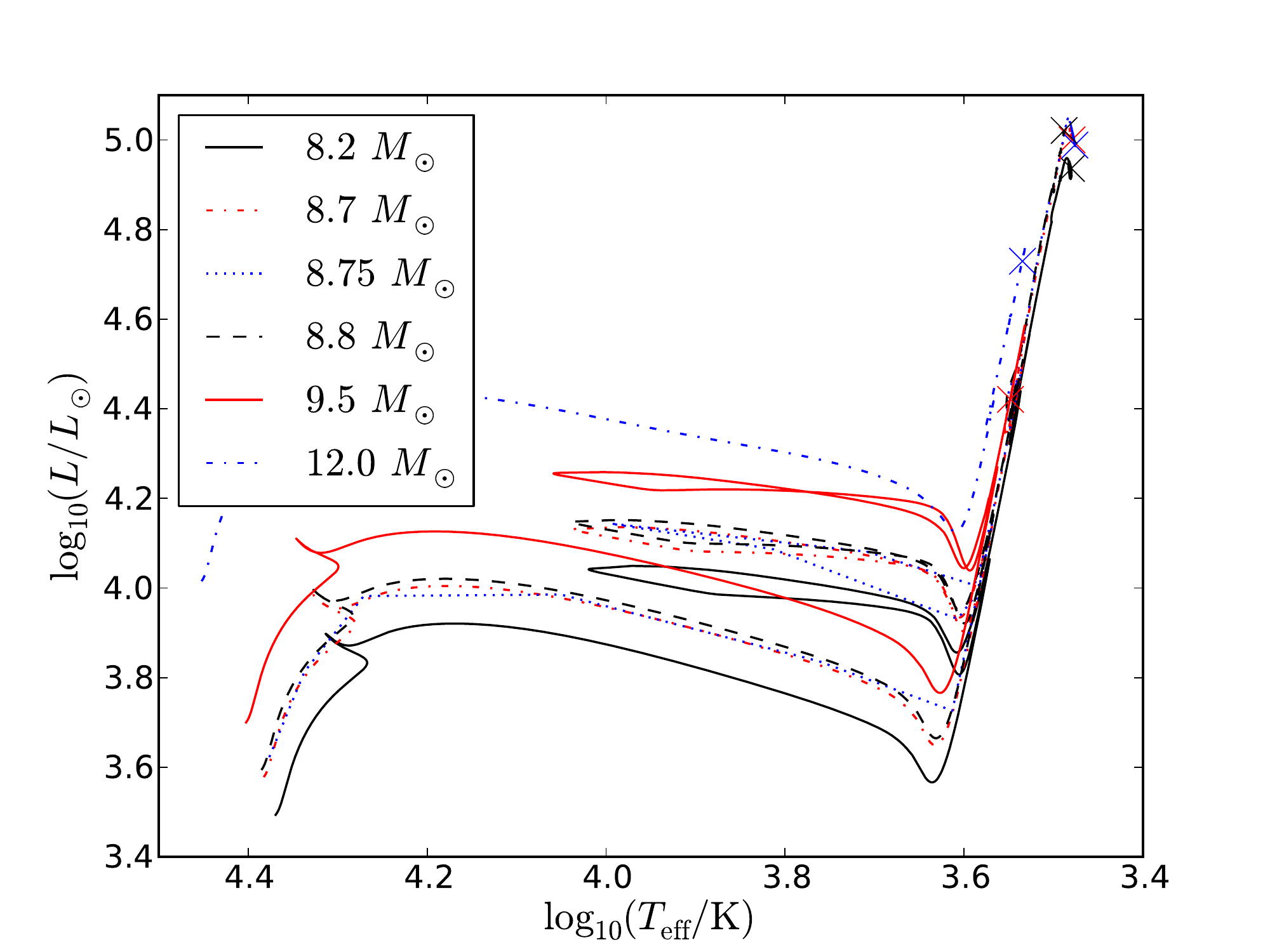}
  \caption{Evolution of all the models in the
Hertzsprung-Russell diagram. By virtue of their deep second dredge up and
subsequent dredge-out, the 8.2, 8.7, 8.75 and 8.8\,M$_\odot$ stars become much more luminous than the
9.5\,M$_\odot$ and even the 12\,M$_\odot$ stars during the late stages. Final luminosities are indicated by crosses.}
  \label{HRD}
\end{figure*}

\begin{figure*}
  \centering
  \subfigure{
  \includegraphics[height=0.47\textwidth,angle=90]{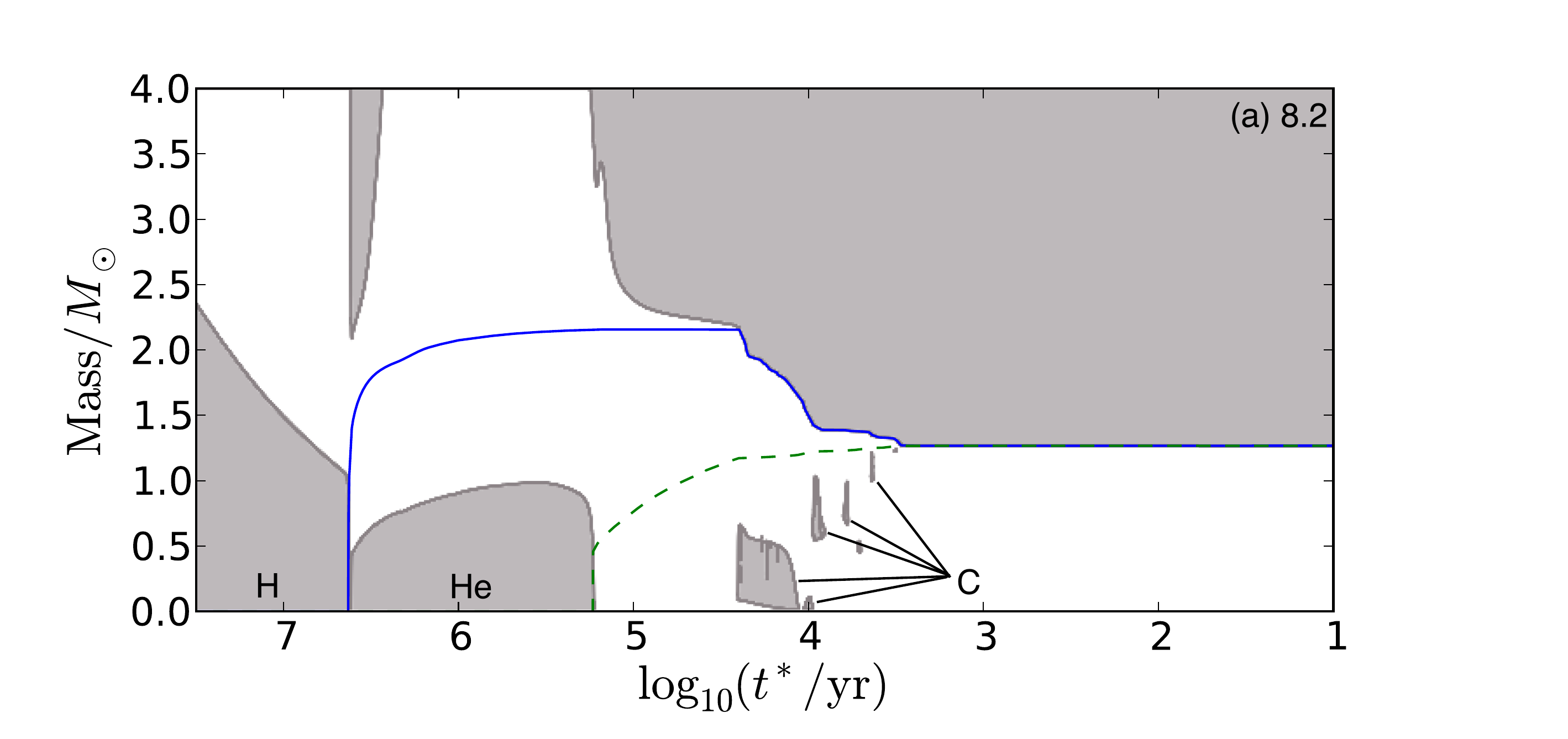}
  \label{kip_8.2}
  }
  \subfigure{
  \includegraphics[height=0.47\textwidth,angle=90]{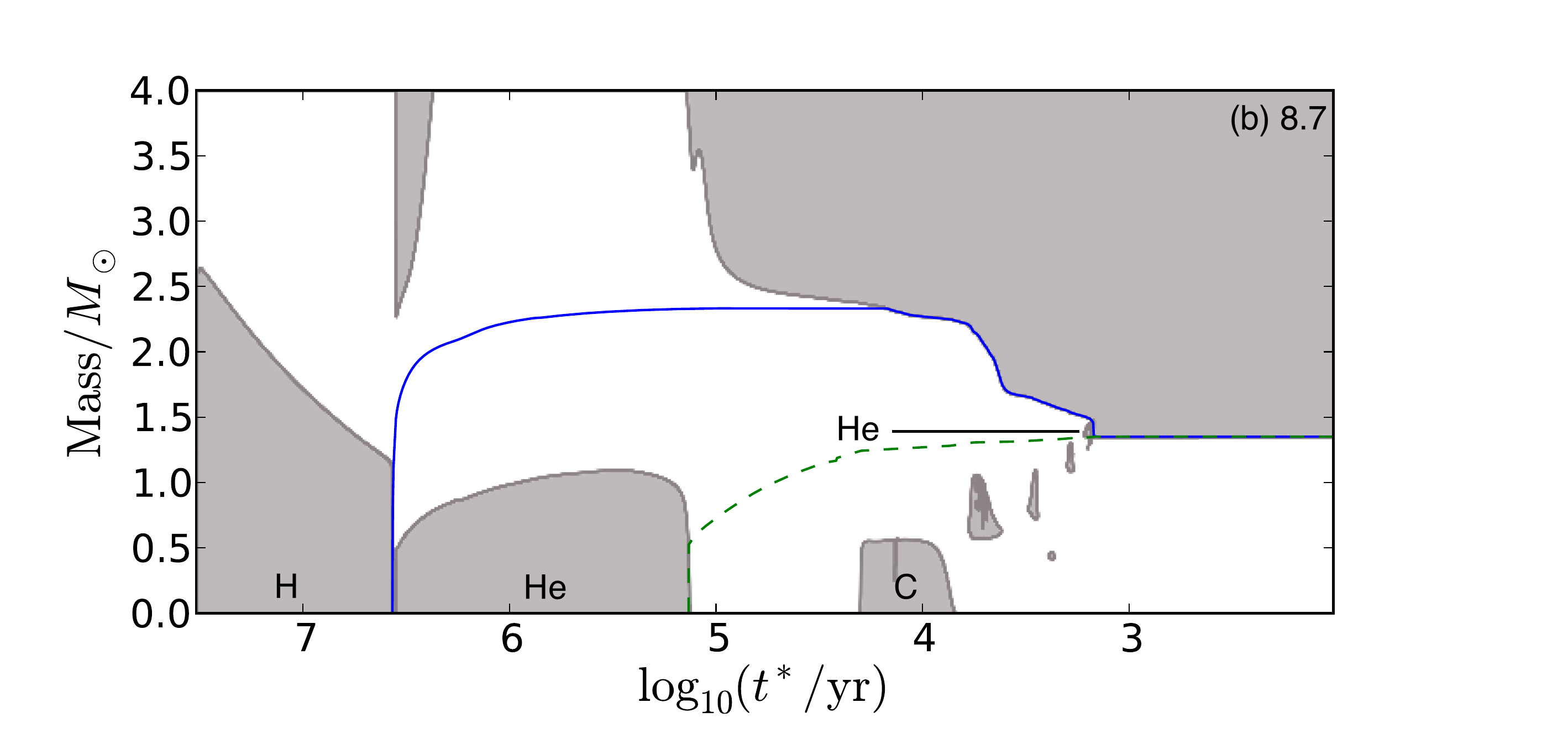}
  \label{kip_8.7}
  }
  \subfigure{
  \includegraphics[height=0.47\textwidth,angle=90]{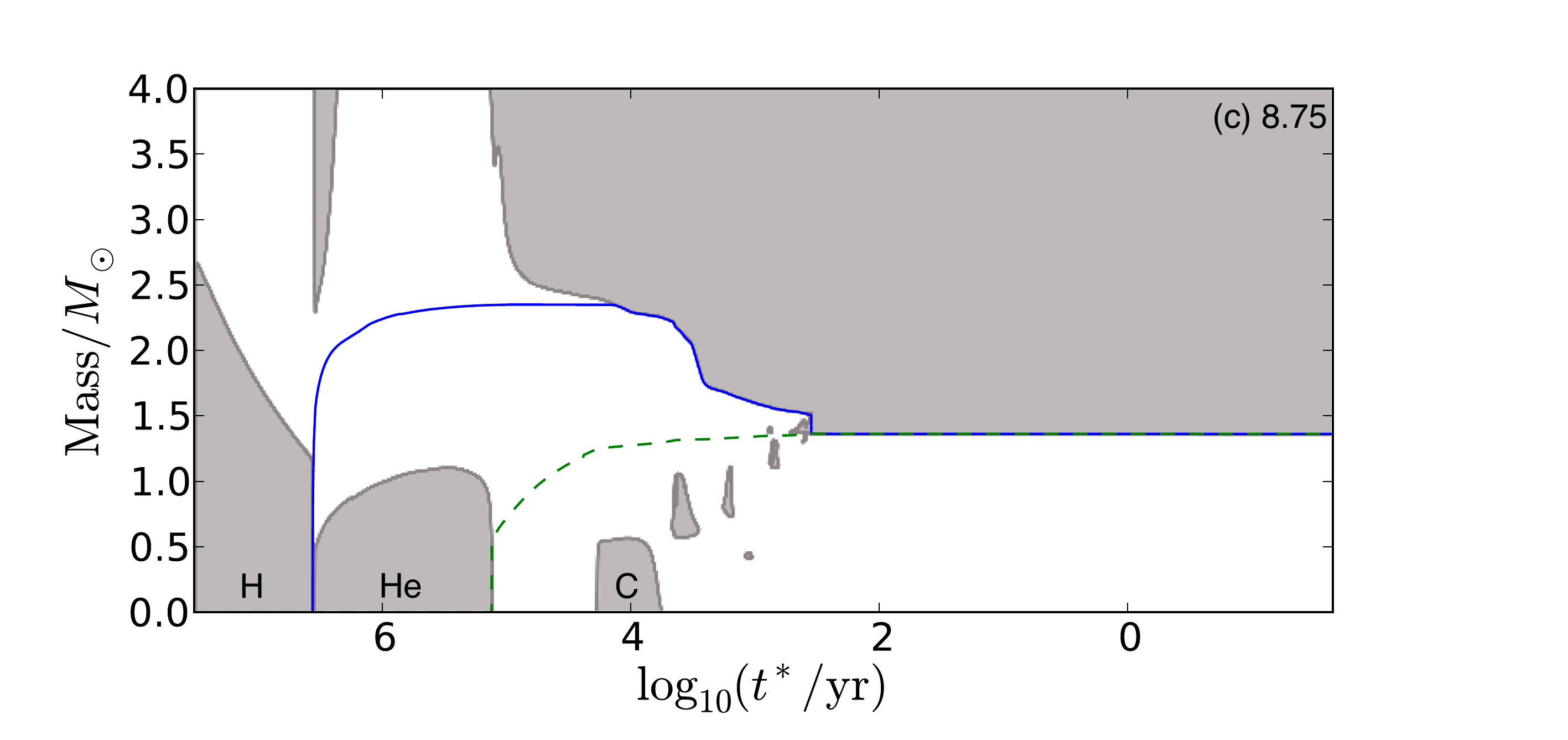}
  \label{kip_8.75}
  }
  \subfigure{
  \includegraphics[height=0.47\textwidth,angle=90]{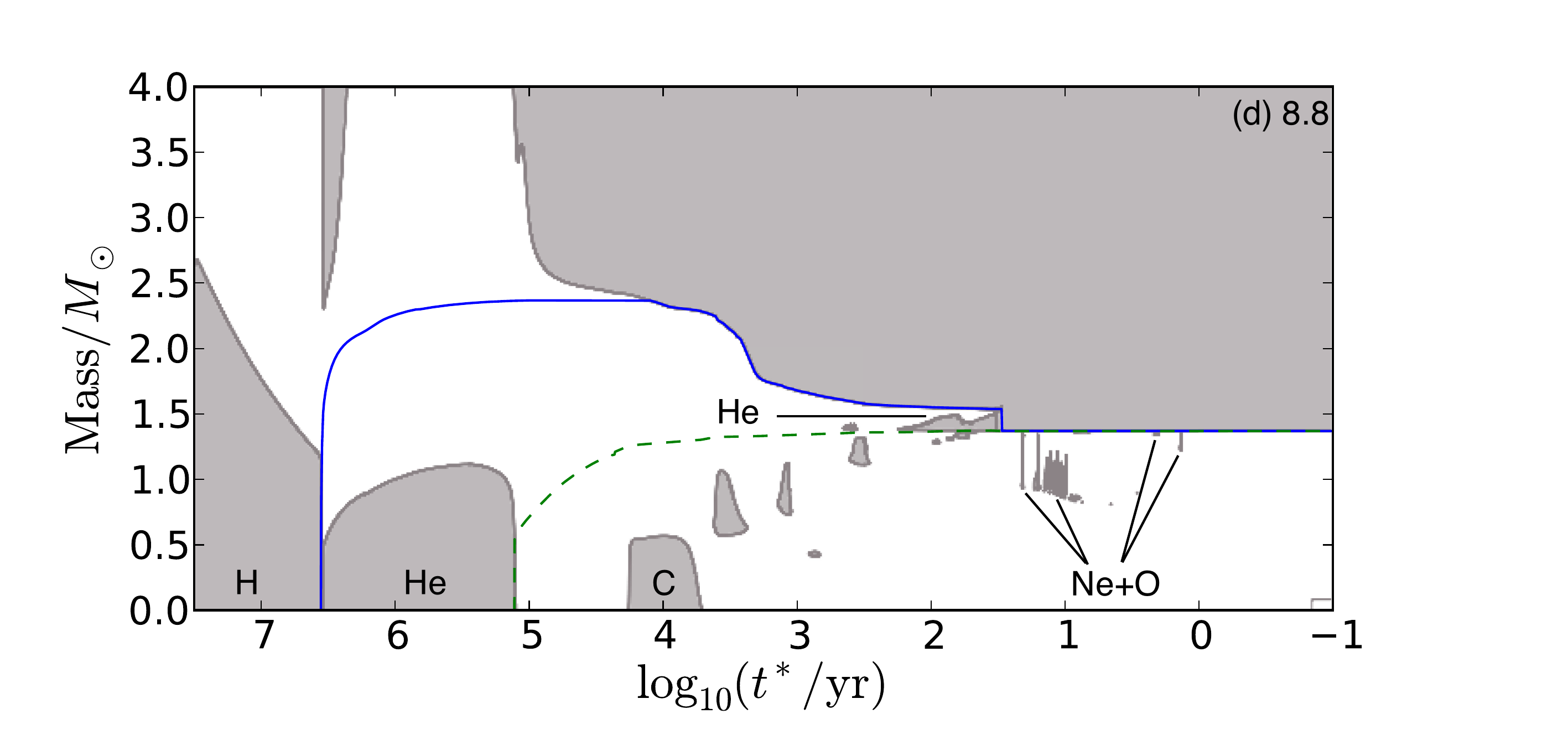}
  \label{kip_8.8}
  }
  \subfigure{
  \includegraphics[height=0.47\textwidth,angle=90]{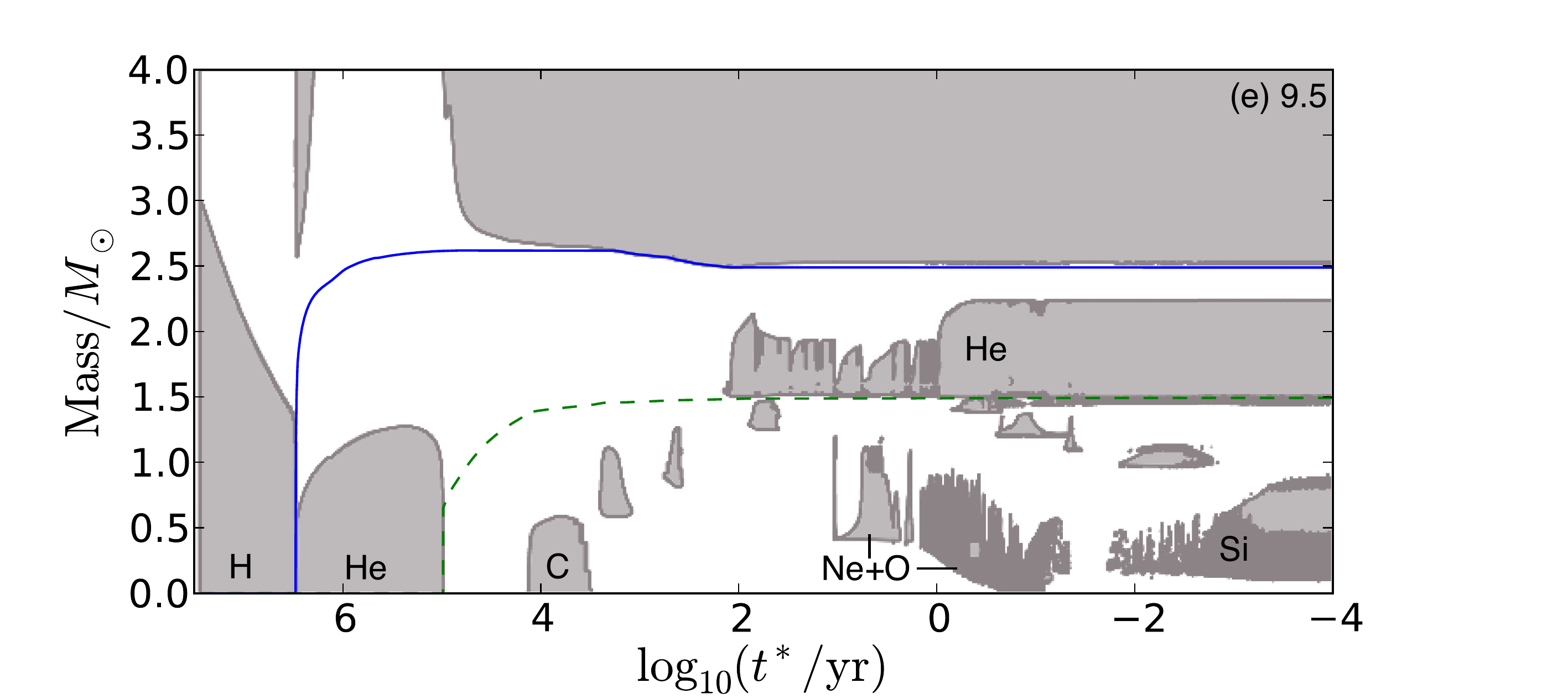}
  \label{kip_9.5}
  }
  \subfigure{
  \includegraphics[height=0.47\textwidth,angle=90]{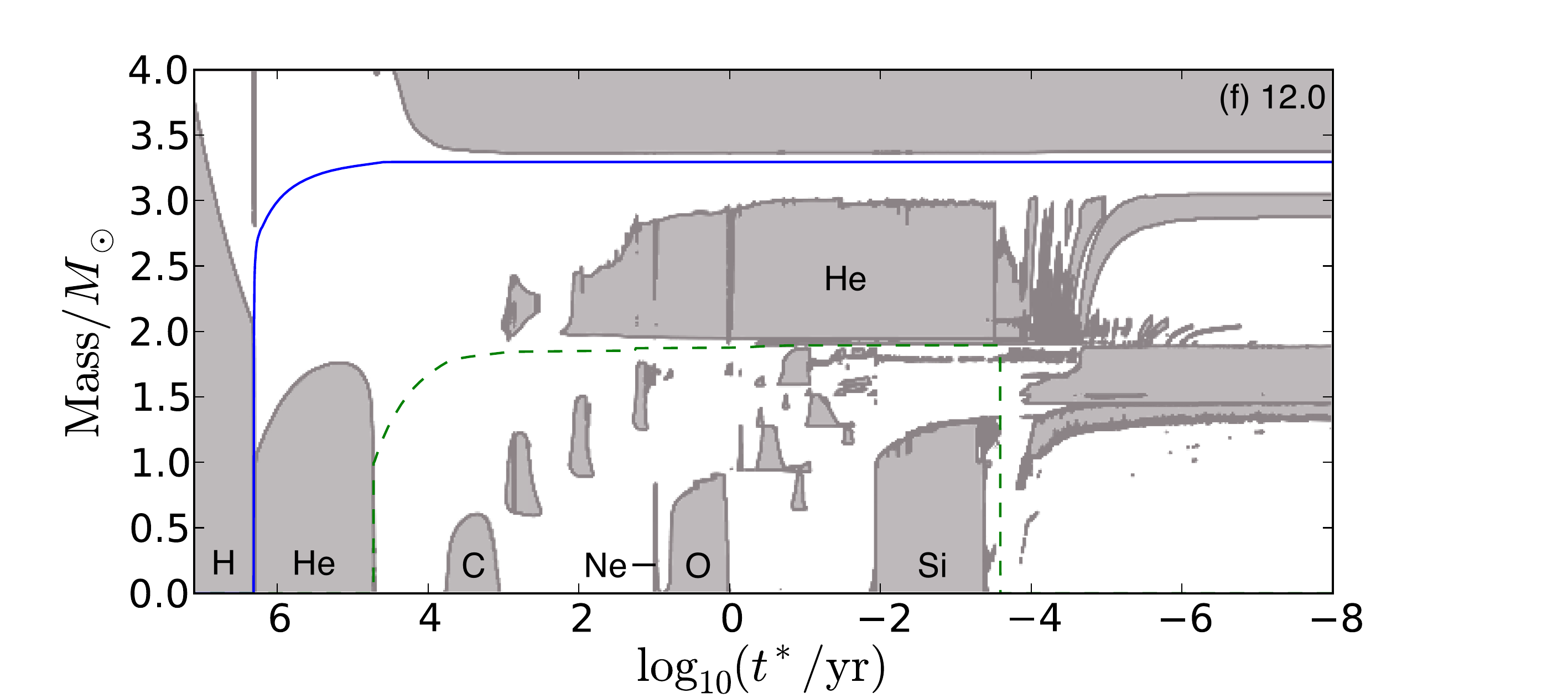}
  \label{kip_12.0}
  }
  \caption{Evolution of convective structure (solid grey shapes) for the H-, He- and advanced burning phases of the models. $t^*/\mathrm{yr}$ is the time left until the end of the calculation. Solid blue and dashed green lines show the locations of the He- (H-free) and CO- (He-free) core boundaries respectively.}
  \label{kips}
\end{figure*}

The evolution of all the models in the Hertzsprung-Russell diagram and the $\rho_\mathrm{c}-T_\mathrm{c}$ plane are shown in Figs.\,\ref{HRD} and \ref{tcrhoc_all_models}, respectively.
Carbon is ignited centrally in all but the $8.2\,M_\odot$ model, in which it is ignited at a mass coordinate of $0.15\,M_\odot$ away from the center and 
the C-burning front propagates to the center (see Fig.\,\ref{kips}a) in the manner of a canonical SAGB flame \citep{Nomoto1984}.

Following the core He-burning stage, core contraction is accompanied by an expansion of the envelope seen in Fig.\,\ref{kips} as a deepening of the base of the convective envelope in mass. Core contraction and the related envelope expansion continues until they are halted by the ignition of carbon. After the exhaustion of carbon in the center, carbon burning proceeds in shells and from this point onwards the behavior of the envelope begins to diverge across the $8-12\,M_\odot$ mass range.

In our models with $M\leq8.8\,M_\odot$ the timescale for expansion of the H-envelope is comparable to the evolutionary timescale. Owing to the higher degree of degeneracy in the core, the envelope in the $8.2\,M_\odot$ model has time to engulf almost the entire helium shell, whereas in the $8.7$, $8.75$ and $8.8\,M_\odot$ models gravo-thermal energy release induces convection in the helium shell that merges with the envelope, referred to as dredge-\emph{out} \citep{Iben1997,Siess2007}. In the $8.8\,M_\odot$ model, as much as $0.8\,M_\odot$ of He-rich material is mixed into the envelope. Aside from the huge increase in the amount of helium that now resides at the surface following this deep mixing event, there are many other observable quantities resulting from dredge-out. In particular, the dredge-out is accompanied by a large increase in luminosity, inducing luminosities at the pre-SN stage larger than for the $12\,M_\odot$ model as shown in Fig.\,\ref{HRD} \citep[see also][]{EldridgeTout2004}.

In the $12\,M_\odot$ model, the evolution of the core is accelerated by plasma neutrino energy losses whereas the envelope expands on a thermal timescale. As a result the convective envelope remains unaltered after carbon burning.
With decreasing initial mass, the core is more degenerate and compact following carbon burning and thus contraction is slower. This provides further energy and time for the expansion of the envelope,
as can be seen at log$_{10}(t^*/\mathrm{yr})\approx4-3$ in Fig.\,\ref{kips}a-e.

\subsection{Late evolution of the $8.2$, $8.7$ and $8.75\,M_\odot$ (super-AGB) models}
\label{SAGBsection}

The $8.2\,M_\odot$, $8.7\,M_\odot$ and $8.75\,M_\odot$ models develop cores with masses that fall short of the critical mass for neon ignition (see \S\,\ref{massivestarsection}) following $2^\mathrm{nd}$ dredge-up ($M_\mathrm{CO} = 1.2670$, $1.3509$ and $1.3621\,M_\odot$ respectively), developing thin (of the order of $10^{-5}-10^{-4}\,M_\odot$) He shells that soon develop a recurrent thermal instability producing transient He-fuelled convection zones (thermal pulses, TP). The $8.2\,M_\odot$ star expels its envelope to become an ONe white dwarf (WD).
It is uncertain whether the $8.7\,M_\odot$ star would produce an ONe WD like the $8.2\,M_\odot$ star, or whether its core would reach the critical central density for electron captures on $^{24}\mathrm{Mg}$, $\rho\approx10^{9.6}\mathrm{\,g\,cm}^{-3}$, before the envelope is lost. We have modelled the $8.75\,M_\odot$ star through the entire TP-SAGB phase (about $2.6\times10^6$ time steps) including the URCA process and electron captures by $^{24}\mathrm{Mg}$ and $^{20}\mathrm{Ne}$ (see Fig.\,\ref{tcrhoc_all_models}). It becomes an EC-SN.

The outcome of these models is highly sensitive to the mass--loss prescription on the SAGB and the rate at which the core grows \citep{Poelarends2008}. We have modeled the TP-SAGB phase of the $8.7\,M_\odot$ star for about 240 pulses, at which point $\rho_\mathrm{c}=10^{9.34}\,\mathrm{g\,cm}^{-3}$. Though still far from $\rho_\mathrm{crit}(^{24}\mathrm{Mg}+\mathrm{e}^-)$, the central density has exceeded the thresholds for both major URCA process reactions, accelerating the contraction of the core towards $\rho_\mathrm{crit}(^{24}\mathrm{Mg}+\mathrm{e}^-)$. Due to this acceleration in contraction and comparison with literature \citep{Nomoto1984,Nomoto1987,Ritossa1999,Poelarends2008}, the most probable outcome for the $8.7\,M_\odot$ model is an EC-SN.

In order to maintain numerical stability in the $8.75\,M_\odot$ model, after the depletion of $^{24}\mathrm{Mg}$ at the center by electron captures, the input physics assumptions were modified. First, the effects of mass--loss were excluded from the calculation and secondly the surface was relocated to a region where the optical depth is an order of magnitude greater than that at the photosphere (which is where the surface had previously been defined).
Choosing to set the boundary at a larger optical depth is one way to deal with the inappropriate way we are simulating the final stages of these massive super-AGB envelopes. In a 1-D code (and probably in the real star) large pulsations occur signalling an increasing instability of the envelope which may lead to enhanced mass loss or even ejection phases, such as the super-wind. These issues have been alluded to recently by \citet{Lau2012}. Choosing the photosphere to be at a larger optical depth indeed lets the star be hotter and smaller, and the mass loss calculated from the stellar parameters, if it were still included, will not be the same as for the default photosphere parameters. Through this treatment, the details of the envelope evolution are increasingly inaccurate from this point. When these changes were made, the remaining envelope mass was $4.48\,M_\odot$ and the central density $\rho_\mathrm{c}=4.67\times10^9\,\mathrm{g\,cm}^{-3}$. For further discussion of numerical instabilities and their physical interpretation, we refer the reader to \citet{WagenhuberWeissRecomb1994} and \citet{Lau2012}.
A simple calculation involving the mass of the envelope at the first thermal pulse of the $8.75\,M_\odot$ model (see Table\,\ref{table_allmodelprops}) and the time spent on the TP-SAGB yields a critical mass--loss rate of $\dot{M}_\mathrm{crit}=6.75\times10^{-4}\,M_\odot\,\mathrm{yr}^{-1}$. That is to say, a mass--loss rate higher than $\dot{M}_\mathrm{crit}$ would have reduced the star to an ONe WD before it could produce an EC-SN. This critical mass--loss rate is within the wide realms applied to super-AGB stars \citep[see][and references therein]{Poelarends2008}.

In contrast to the $8.8\,M_\odot$ model, which is discussed in \S\,\ref{massivestarsection}, there is no significant $Y_\mathrm{e}$ reduction in the outer core, since there was no Ne-O flash. Instead, the contraction is driven by the steady growth of the core during each thermal pulse and the contraction is slower and heating competes with neutrino losses so that the core resumes cooling until electron captures by $^{24}\mathrm{Mg}$ are activated (see Fig.\,\ref{tcrhoc_all_models}). The difference can again be seen following the depletion of $^{24}\mathrm{Mg}$ at the center of both models, where the $8.8\,M_\odot$ model continues to heat while the $8.75\,M_\odot$ model again cools down. This difference in temperature between the center of the $8.8\,M_\odot$ and $8.75\,M_\odot$ models is important when considering the next phase of their evolution - electron captures by $^{20}\mathrm{Ne}$.

\subsection{Late evolution of the $8.8$, $9.5$ and $12.0\,M_\odot$ (massive star) models}
\label{massivestarsection}
The mass of the CO core, $M_\mathrm{CO}$, continues to grow for the entire lifetime of the secondary C-burning shells in all models due
to helium shell burning. Previous studies \citep[see][and references therein]{Nomoto1984} show that the core mass limit for neon ignition is very close to $1.37\,M_\odot$, which our models confirm. Indeed, in all models with initial mass greater than $8.8\,M_\odot$, a CO-core develops, with a mass that exceeds the limit for neon ignition, $M_\mathrm{CO}(8.8\,M_\odot,\,9.5\,M_\odot,\,12.0\,M_\odot)=1.3696,\,1.4925,\,1.8860\,M_\odot$.

\begin{figure*}
  \centering
  \includegraphics[width=0.44\textheight,angle=90]{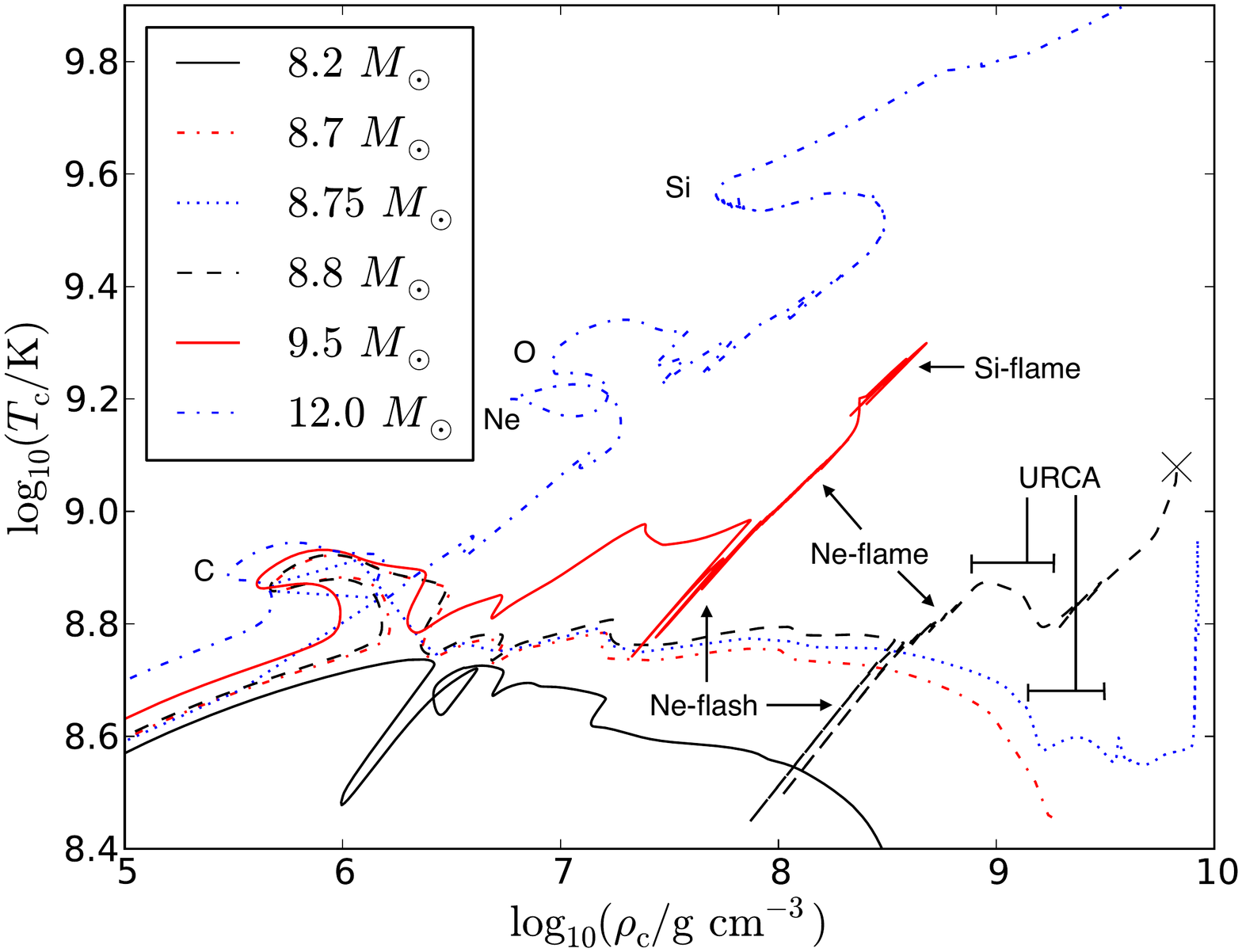}
  \caption{The divergence of the models following C-burning in the $\mathrm{log}_{10}(\rho_\mathrm{c})-\mathrm{log}_{10}(T_\mathrm{c})$ plane; the cross shows from where the evolution of the 8.8\,M$_\odot$ model was continued with the AGILE-BOLTZTRAN 1-D hydro-code.}
  \label{tcrhoc_all_models}
\end{figure*}

\begin{figure*}
  \centering
  \includegraphics[height=0.35\textheight]{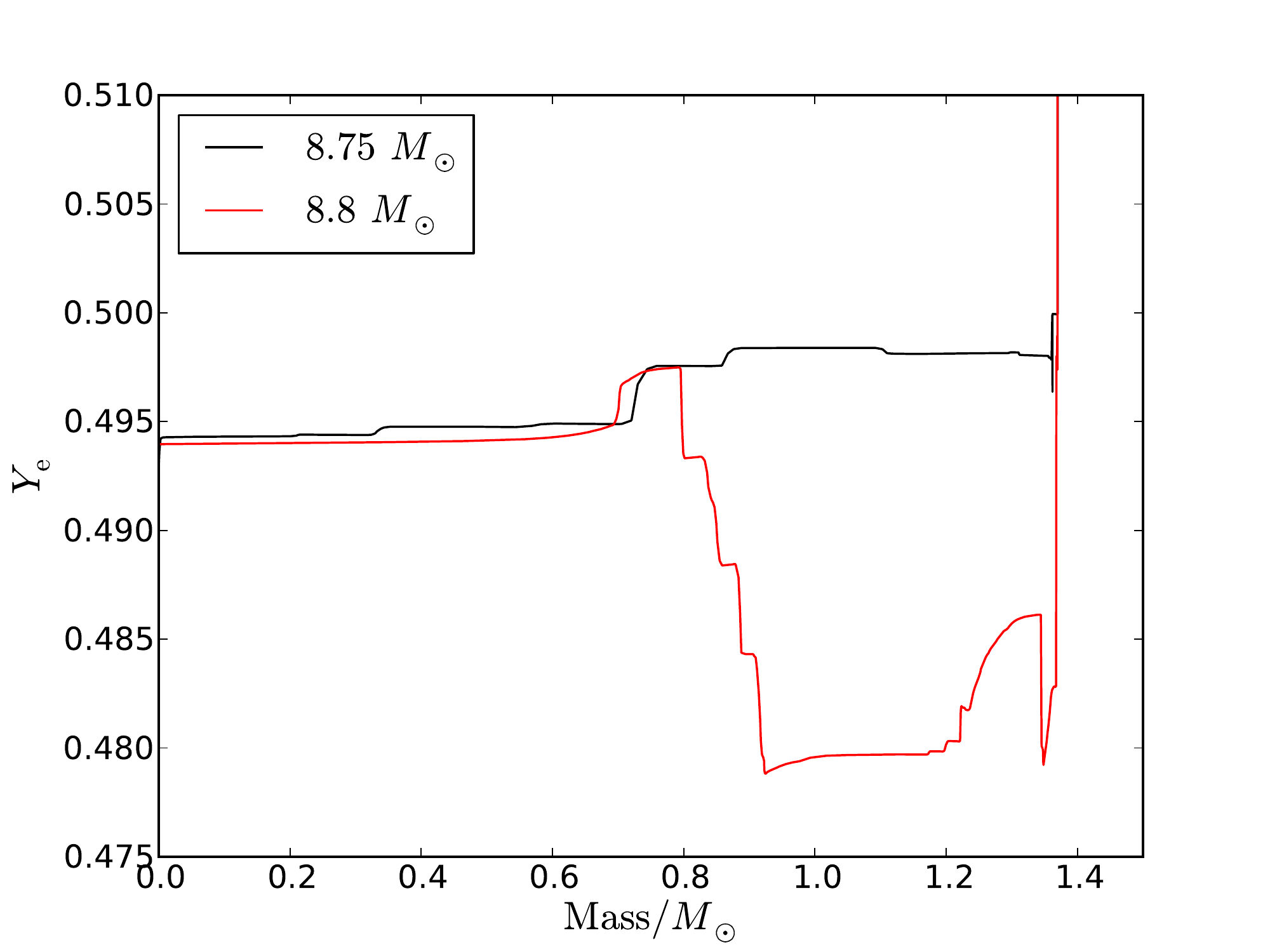}
  \caption{Radial profiles of the electron fraction, $Y_\mathrm{e}$, in the progenitor structure of the 8.8\,M$_\odot$ model and the 8.75\,M$_\odot$ model after central $^{24}\mathrm{Mg}$ depletion. The silicon-rich shell of the 8.8\,M$_\odot$ model, where the material has been processed by the neon-oxygen shell flashes, displays a severely reduced electron fraction, reflected by high ratios of $^{30}\mathrm{Si}/^{28}\mathrm{Si}$ and $^{34}\mathrm{S}/^{32}\mathrm{S}$.}
  \label{yeprofs}
\end{figure*}

A temperature inversion develops in the core following the extinction of carbon-burning in both the $8.8\,M_\odot$ and $9.5\,M_\odot$ models. The neutrino emission processes that remove energy from the core are (over-) compensated by heating from gravitational contraction in more massive stars. However in these lower-mass stars the onset of partial degeneracy moderates the rate of contraction and hence neutrino losses dominate, cooling the central region. As a result, the ignition of neon in the 8.8 and 9.5~$M_\odot$ models takes place off center, at mass coordinates of $0.93\,M_\odot$ and $0.40\,M_\odot$ respectively. This result confirms the work of \citet{Nomoto1984} (case 2.6), but diverges from that of \citet{EldridgeTout2004}, which we will discuss later. In both models the temperature in the neon-burning shell becomes high enough to also ignite $^{16}\mathrm{O}\,+\,^{16}\mathrm{O}$. As we mention in \S\,\ref{methodsec}, owing to the high densities in the cores of these stars, the products of neon and oxygen burning are more neutron-rich than in more massive stars. This results in an electron fraction in the shell of as low as $Y_\mathrm{e}\approx0.48$ (see \S\,\ref{progenitorstructures} and Fig.\,\ref{yeprofs}). Such low $Y_\mathrm{e}$ causes the adiabatic contraction in the following way. If the temperature is high during the flash, the flashing outer layer expands and exerts lower pressure (less weight) on the central region (as can be seen in Fig.\,\ref{tcrhoc_all_models} labelled `Ne-flash', $\rho_\mathrm{c}$ decreases due to the almost adiabatic expansion of the central region). However, when the flashed region has cooled down by neutrino emission following the extinction of nuclear burning, the outer layer shrinks and exerts more weight on the core, which is less able to provide support than before the flash because there are fewer electrons available to contribute to the degeneracy pressure. The center then reaches higher densities, and hence temperatures, than before. As mentioned above, for this reason the reduction in $Y_\mathrm{e}$ is important for cores so close to $M_\mathrm{Ch}$.

As illustrated in Fig.\,\ref{kips}e, following the neon shell flashes the $9.5\,M_\odot$ model recurrently ignites neon- and oxygen-burning in shells at successively lower mass coordinates that eventually reach the center, following which Si-burning is ignited off-center.
Although neon- (and oxygen-) burning in the $8.8\,M_\odot$ model begins as a flash and later propagates toward the center, the evolution of the $8.8\,M_\odot$ model diverges from that of the $9.5\,M_\odot$ star when its center reaches the conditions necessary for the first URCA process pair to become significant (whereas the $9.5\,M_\odot$ model avoids such dense conditions). More details of the neon and oxygen shell burning episodes are discussed in \S\S\,\ref{lateev} and \ref{lateev2}.

The CO core (or equivalently He-free core) in the $8.8M_\odot$ model at the time of neon ignition is $1.36964\,M_\odot$, very close to $M_\mathrm{Ch}$, while that of the $9.5\,M_\odot$ model is $1.49246\,M_\odot$ (see Table\,\ref{table_allmodelprops}). Under these conditions, the $8.8\,M_\odot$ model experiences a much more marked contraction due to the reduction in $Y_\mathrm{e}$. The central density at this time is as high as $3.43\times10^8\mathrm{\,g\,cm}^{-3}$, which is exceedingly close to the threshold density for $^{27}\mathrm{Al}(\mathrm{e}^-,\nu)^{27}\mathrm{Mg}$. Although there is no cooling effect from the A=27 pair because the decay channels are blocked, the further removal of electrons from the core causes contraction toward the threshold densities of the second and third URCA pairs (A=25 and A=23 respectively). The cooling effect supplied by the A=25 URCA pair (and later the A=23 pair, shown in Fig.\,\ref{tcrhoc_all_models}) allows for a small amount of contraction but again it is the associated change in the electron fraction that enables the largest contraction when the core is so close to the Chandrasekhar limit ($M_\mathrm{Ch}\propto Y_\mathrm{e}^2$). The core of the $8.8\,M_\odot$ model continuously contracts until the center reaches
the critical density for electron captures by $^{24}\mathrm{Mg}$, quickly
followed by further contraction to the critical density for those by
$^{20}\mathrm{Ne}$ (see Fig.\,\ref{tcrhoc_all_models}).

\begin{figure*}
  \centering
  \includegraphics[width=0.7\textwidth]{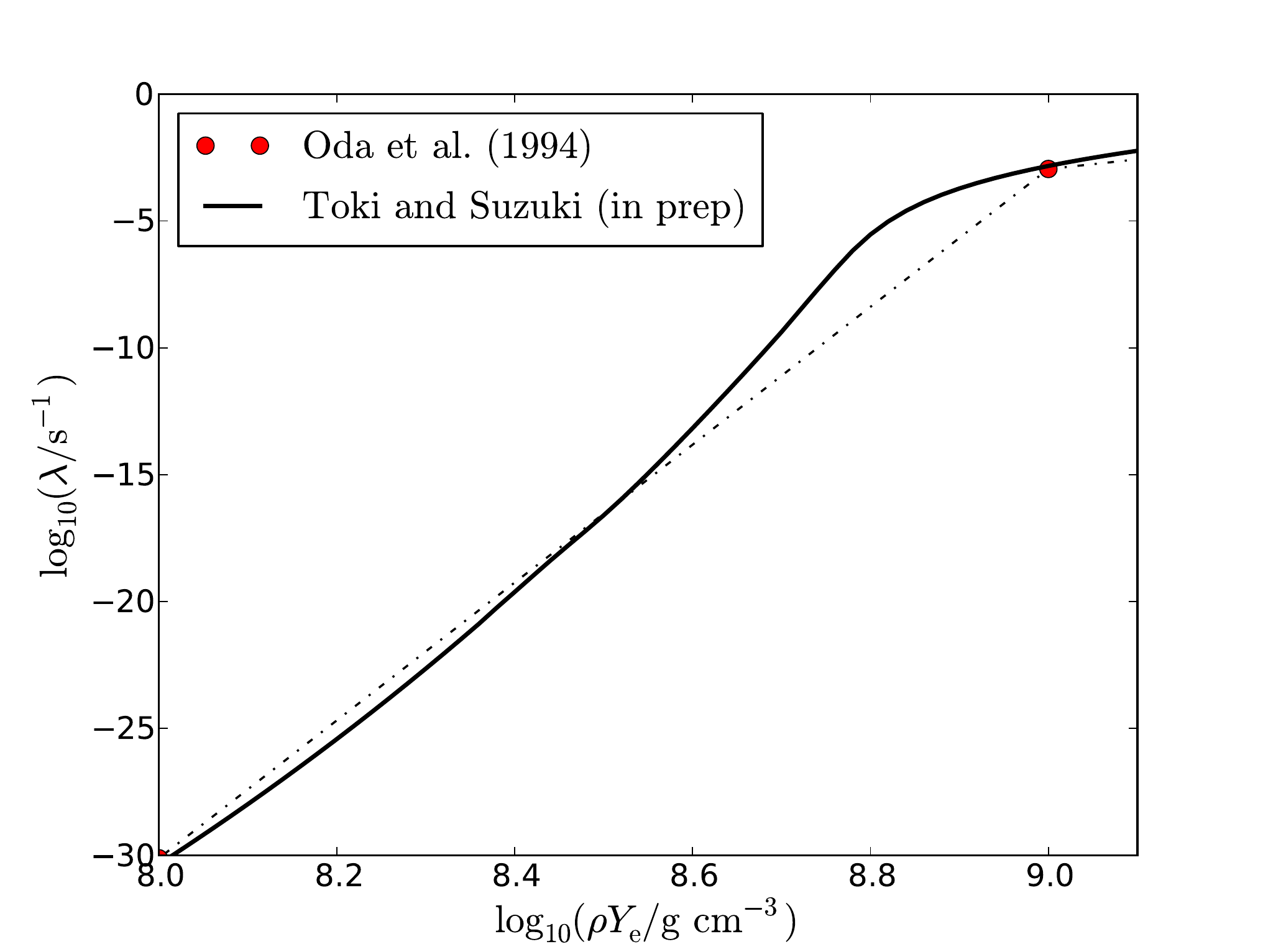}
  \includegraphics[width=0.7\textwidth]{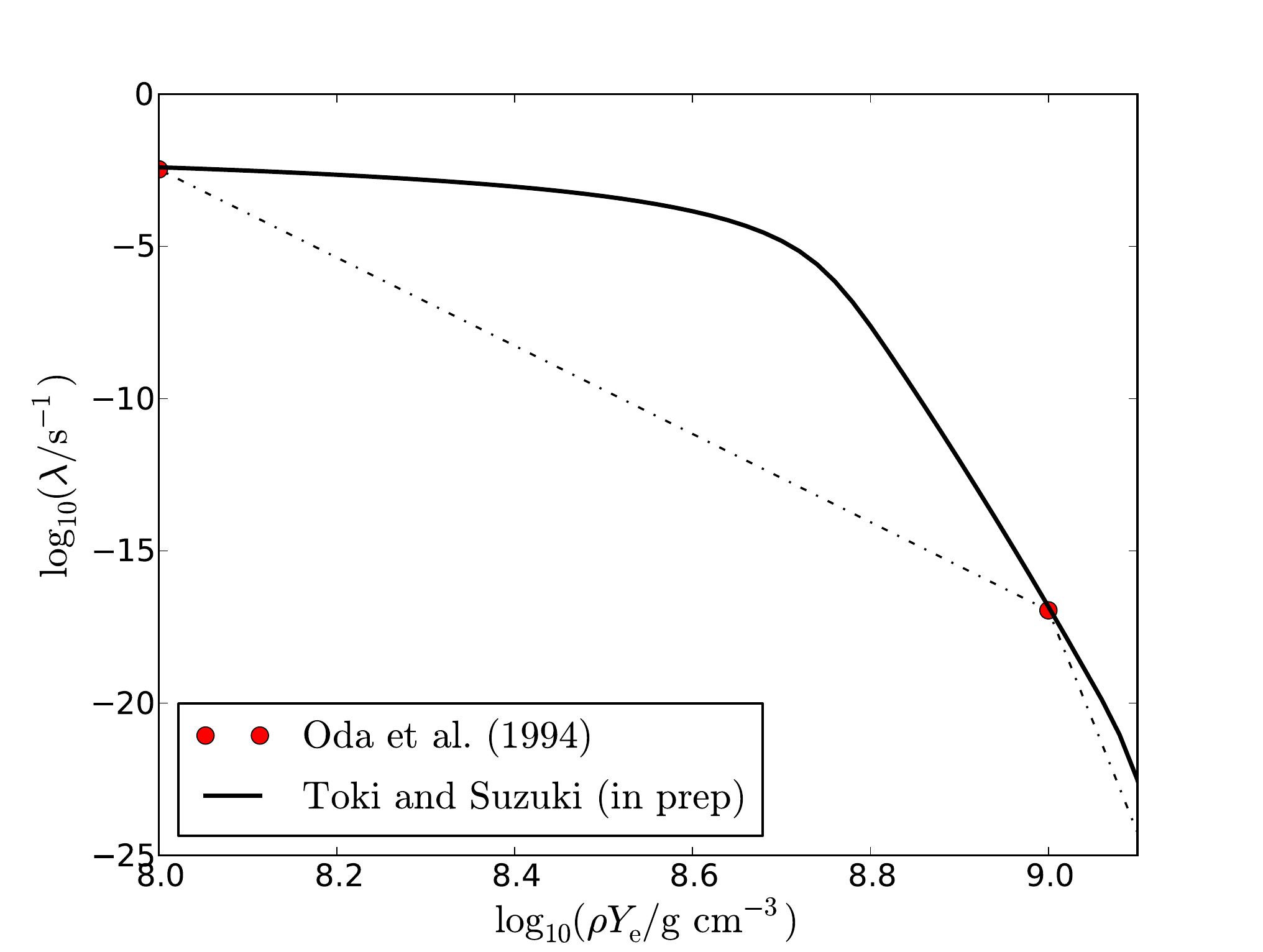}
  \caption{$^{25}\mathrm{Mg}$ electron capture rate (left panel) and $^{25}\mathrm{Na}$ beta decay rate (right panel) at $T=4\times10^8\,\mathrm{K}$ from the compilation of \citet{ODA94} and the new calculation by \citet{Toki2013}.}
  \label{toki_vs_oda_rates}
\end{figure*}

There is a significant discrepancy between the URCA-process trajectories of our models and those of \citet{Ritossa1999}. This is due to the under-sampling of weak reaction rates for the URCA process that we employ in the MESA code \citep{ODA94}. In \S\,\ref{weakdiscussion} we discuss the implications of this under-sampling and show that, by using new well sampled weak rates \citep{Toki2013}, the URCA process central trajectory of \citet{Ritossa1999} is qualitatively reproduced in the $8.8\,M_\odot$ case. The difference between the \citet{ODA94} compilation and the newly calculated \citet{Toki2013} rates that are available to use in our calculations for the A=25 pair is shown in Fig.\,\ref{toki_vs_oda_rates}.

This central evolution is significantly different from that for the $8.75\,M_\odot$ model, which is described in \S\,\ref{SAGBsection}.
The energy release from both the rapid contraction and the
$\gamma$-decays from electron--capture products raise the temperature high enough to ignite neon and oxygen in quick succession. We have followed from the resulting oxygen deflagration onwards with the AGILE-BOLTZTRAN hydrodynamics code and can confirm that the model results in core collapse \citep{FischerECSN2013}.
Although \citet{EldridgeTout2004} report the same fate for their $10\,M_\odot$ model in which a limited network was used, there is no neon shell flash following the completion of the second dredge-up.
In these models, neon burning 
was found to take place at the edge of the core during the last carbon-shell flash, reducing the core mass to $M_\mathrm{Ch}$ \citep[][private communication]{Eldridge_thesis}. Subsequently, the core contracted directly to central densities of about $\mathrm{log}_{10}(\rho_\mathrm{c}/\mathrm{g\,cm}^{-3})=9.8$ (roughly the critical density for electron captures by $^{20}\mathrm{Ne}$ to start) with no further neon-shell flashes, though electron captures were not included in the nuclear reaction network. Neon-burning reaction rates were artificially limited to prevent numerical problems and a low spatial resolution was used. We believe these two caveats to be the reason that the neon-oxygen shell flashes we find to occur in such stars were not present in these earlier models.
In this work we were able to follow the evolution all the way to oxygen deflagration by using a very large network of 114 nuclei including all the relevant fusion and weak reactions. Our models thus highlight the importance of neon-shell burning in determining the path to collapse.

As mentioned above, the $9.5\,M_\odot$ model starts silicon burning off center in a shell that later propagates toward the center. This is another example of the continuous transition towards massive stars, in which all the burning stages begin centrally. Although we have not evolved this model to its conclusion, we expect that silicon-burning will migrate to the center, producing an iron core, and that it will finally collapse as an FeCCSN. Such a low--mass progenitor will make for interesting explosion simulations \citep{Mueller2012}. 
The $12\,M_\odot$ is the canonical massive star in our grid, igniting C-, Ne-, O- and Si-burning centrally (see Fig.\,\ref{kips}f). It eventually collapses, and would produce a type-II FeCCSN.

\subsubsection{Neon-oxygen flashes}
\label{lateev}
As briefly mentioned above, following the extinction of the final carbon burning shell, a degeneracy/neutrino-induced
temperature inversion arises in the core in a similar way to the temperature
inversion in SAGB stars. This causes the ignition of carbon to take place away
from the center. Neon is thus ignited off-center at mass co-ordinates of
$0.93M_\odot$ and $0.40M_\odot$ for the $8.8M_\odot$ and $9.5M_\odot$ models
respectively. Some of the important model properties are given in
Table\,\ref{table_allmodelprops} at this time.

At the point of Ne-shell ignition, the density profile of the two stars is very
different (see Fig.\,\ref{rhoprof_ne_ignition}). While the $8.8M_\odot$ model is structured more like a super-AGB star
due to the previous dredge-out episode, the $9.5M_\odot$ model resembles more a
massive star, with a distinct He-shell and C-shell still present.

When neon and oxygen is first ignited, fuel is abundant and a convective shell
quickly develops.
The sharp increase in energy production briefly halts the contraction of the
core and causes the center to expand and cool (see Fig.\,\ref{tcrhoc_all_models}),
while convection brings in fresh
fuel to be burnt at the base of the shell.
Conduction at the base of the shell
is slow and so when the fuel is depleted the shell is extinguished and the core
contracts. This contraction continues until the temperature becomes high enough where neon and oxygen is abundant, re-igniting the nuclear burning and producing a new convective shell.
After a few flashes, the region previously engulfed by the shell convective shell
as it extended radially outwards has become heavily depleted in Ne and O and so the closest fuel is in the direction
of the center. At this point a new regime, the Ne-O flame, is begun.

\subsubsection{Neon-oxygen flame}
\label{lateev2}
In this section, and throughout the remainder of the manuscript, it should be noted that we use the term flame to describe the inward propagation of a nuclear burning shell, driven by either compressional heating or other form of local heat transport towards the center.
After the last flash has extinguished and contraction begins, the
two models begin to diverge, as best illustrated in Figs\,\ref{tcrhoc_all_models} and \ref{kips}(d, e). The $9.5M_\odot$ star once again contracts and a
thin shell of neon and oxygen is ignited below the base of the previously
convective shell. Any convection developing at this time does not bring any
fresh fuel (only the ashes of the previous shells) into the burning region. The
core is so dense that the photon mean free path is too short for radiative
transfer to play an important role in the inward propagation of the flame and
instead compressional heating due to core contraction and local heating due
to electron conduction are largely responsible
for intermittent periods of nuclear energy production that move towards the
center.

It is a different story for the $8.8M_\odot$ star. Contraction, following the
final ONe-shell flash, at first acts to heat the material locally and to
burn neon and oxygen moderately as in the
$9.5M_\odot$ model, except that the core is more degenerate in the $8.8\,M_\odot$ star.
Electron conduction is therefore much more efficient and the localised effect of
heat generation due to contraction and any subsequent nuclear burning is instead
diluted across the core. This
smoothing of the temperature profile across the core prevents the region directly below
the previously ONe-burning shells from reaching temperatures in excess
of the Ne-burning threshold. Instead of a flame developing as in the
$9.5M_\odot$ star, the core contraction, driven by the neutron-rich composition in the NeO shell, causes local heating much further from the center where the degeneracy is lower, and a new neon and oxygen burning shell ignites (where the fuel is still abundant) above the outermost extent of the previous ONe-shells.

\begin{figure*}
  \centering
  \subfigure[8.8\,M$_\odot$ early flame]{
  \includegraphics[width=0.45\textwidth]{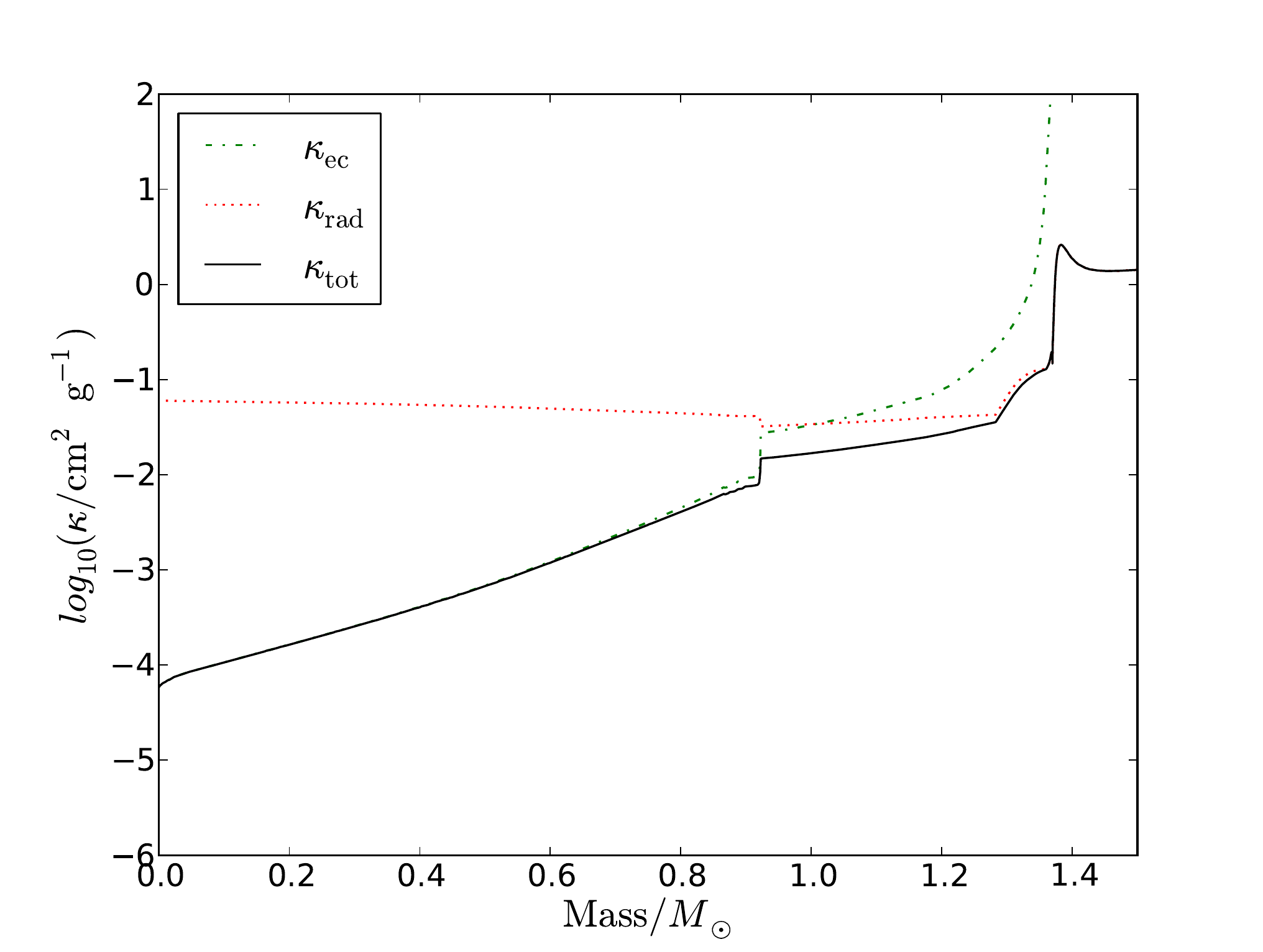}
  \label{kappa_profiles_88_early}
  }
  \subfigure[9.5\,M$_\odot$ early flame]{
  \includegraphics[width=0.45\textwidth]{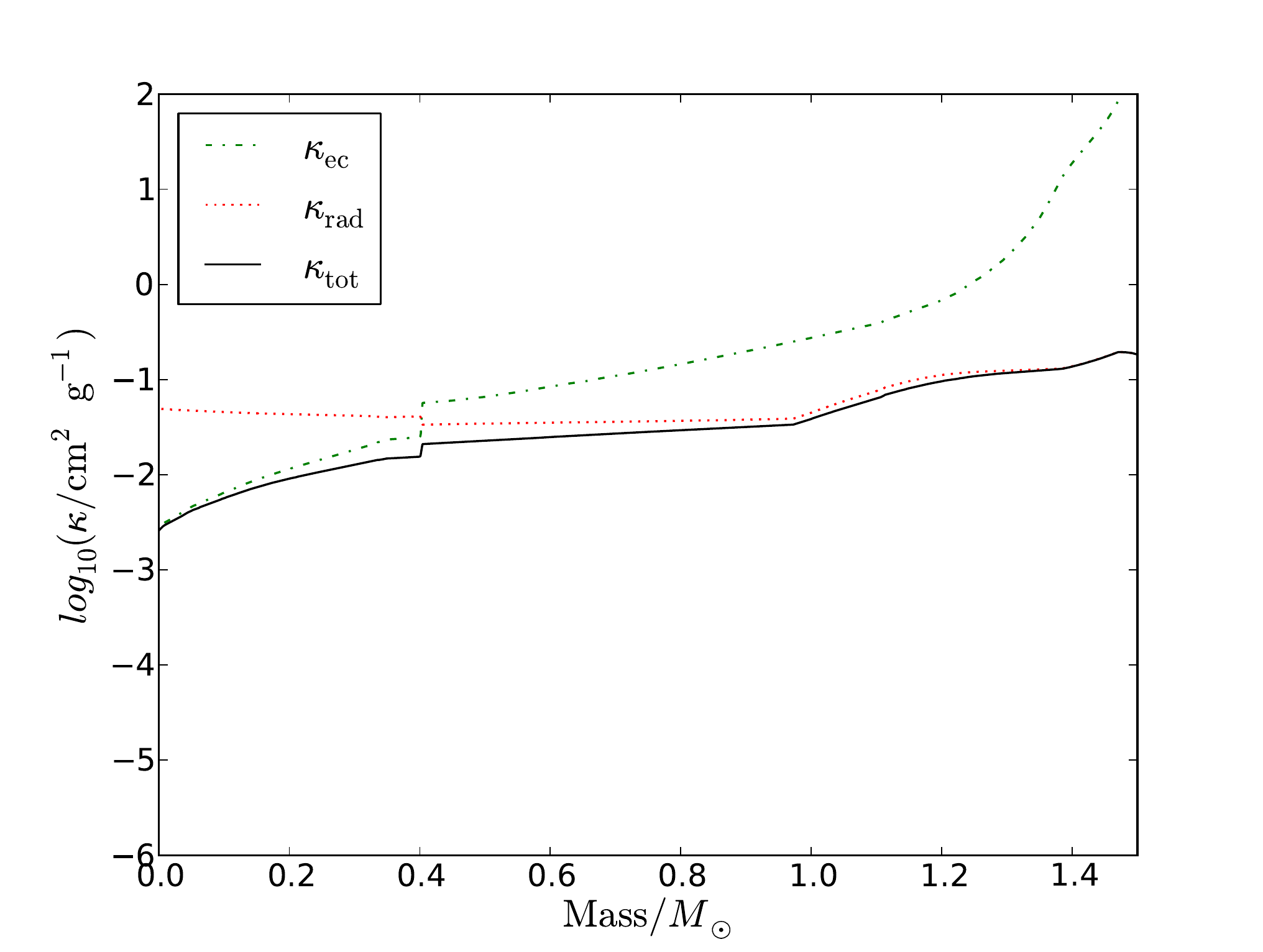}
  \label{kappa_profiles_95_early}
  }
  \subfigure[8.8\,M$_\odot$ flame failure]{
  \includegraphics[width=0.45\textwidth]{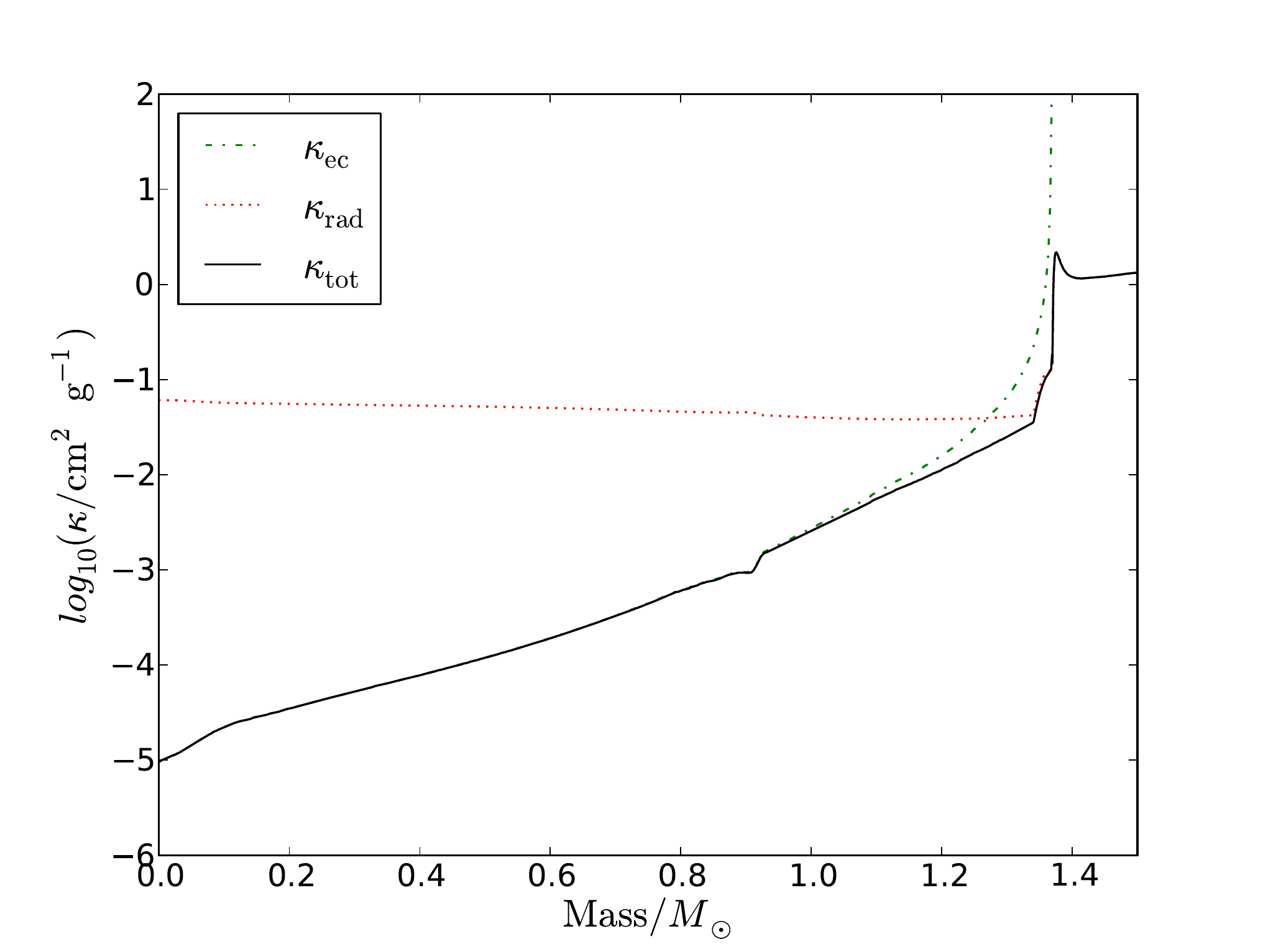}
  \label{kappa_profiles_88_late}
  }
  \subfigure[9.5\,M$_\odot$ late flame]{
  \includegraphics[width=0.45\textwidth]{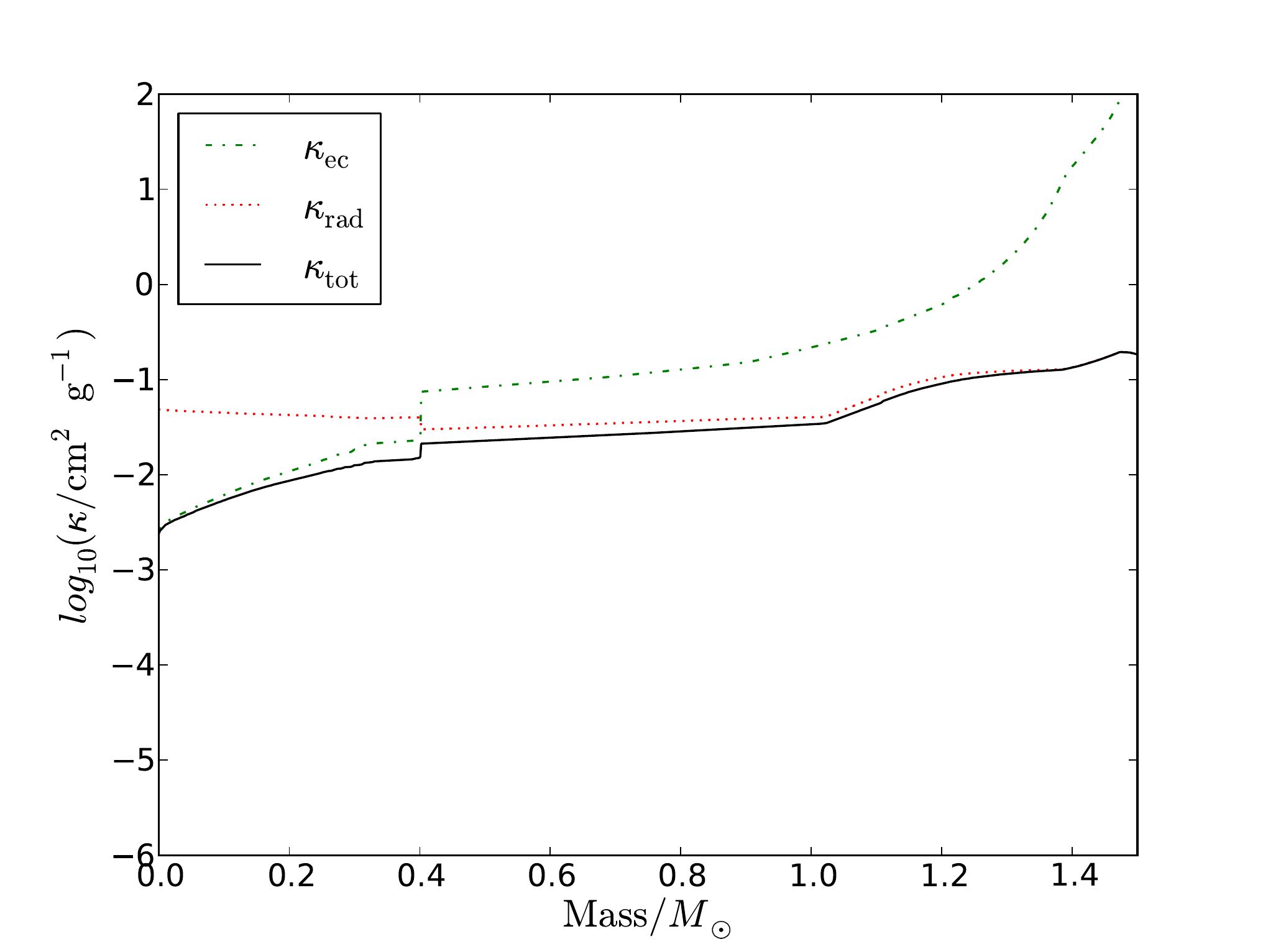}
  \label{kappa_profiles_95_late}
  }
  \caption{Radial profiles with respect to mass co-ordinate of the radiative ($\kappa_\mathrm{rad}$), conductive ($\kappa_\mathrm{ec}$)
and total ($\kappa_\mathrm{tot}=[1/\kappa_\mathrm{rad}+1/\kappa_\mathrm{ec}]^{-1}$) opacities following the extinction of the final
neon-oxygen convective flash episode. The heat transport in both stars is
dominated by conduction (lower $\kappa$), however a stable nuclear flame only develops in the
9.5\,M$_\odot$ model by virtue of its higher total opacity allowing for heating to
take effect on a much more local scale than in the 8.8\,M$_\odot$ model.}
  \label{kappa_profiles}
\end{figure*}

\begin{figure*}
  \centering
  \subfigure[8.8\,M$_\odot$]{
  \includegraphics[width=0.45\textwidth]{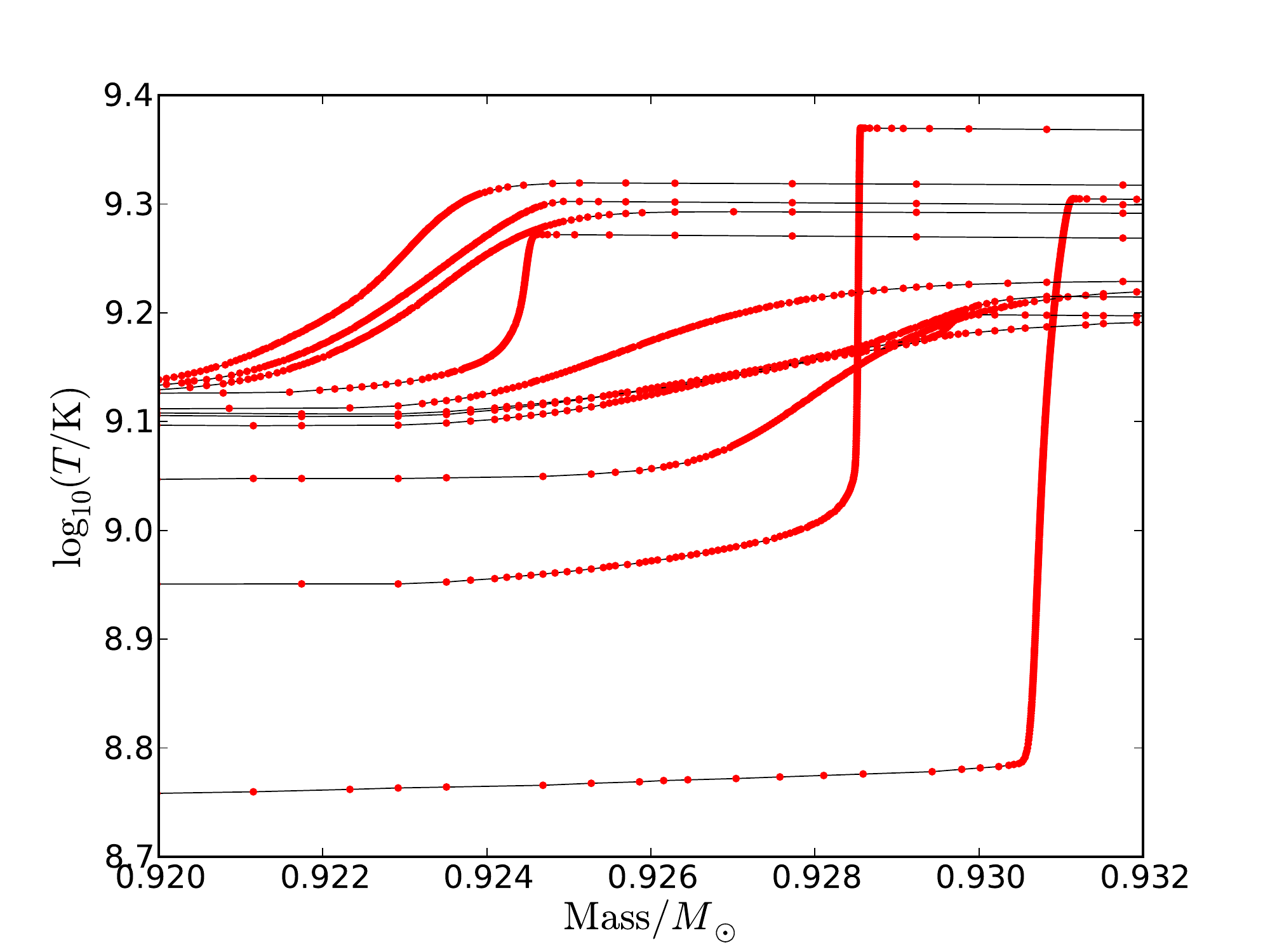}
  \label{}
  }
  \subfigure[9.5\,M$_\odot$]{
  \includegraphics[width=0.45\textwidth]{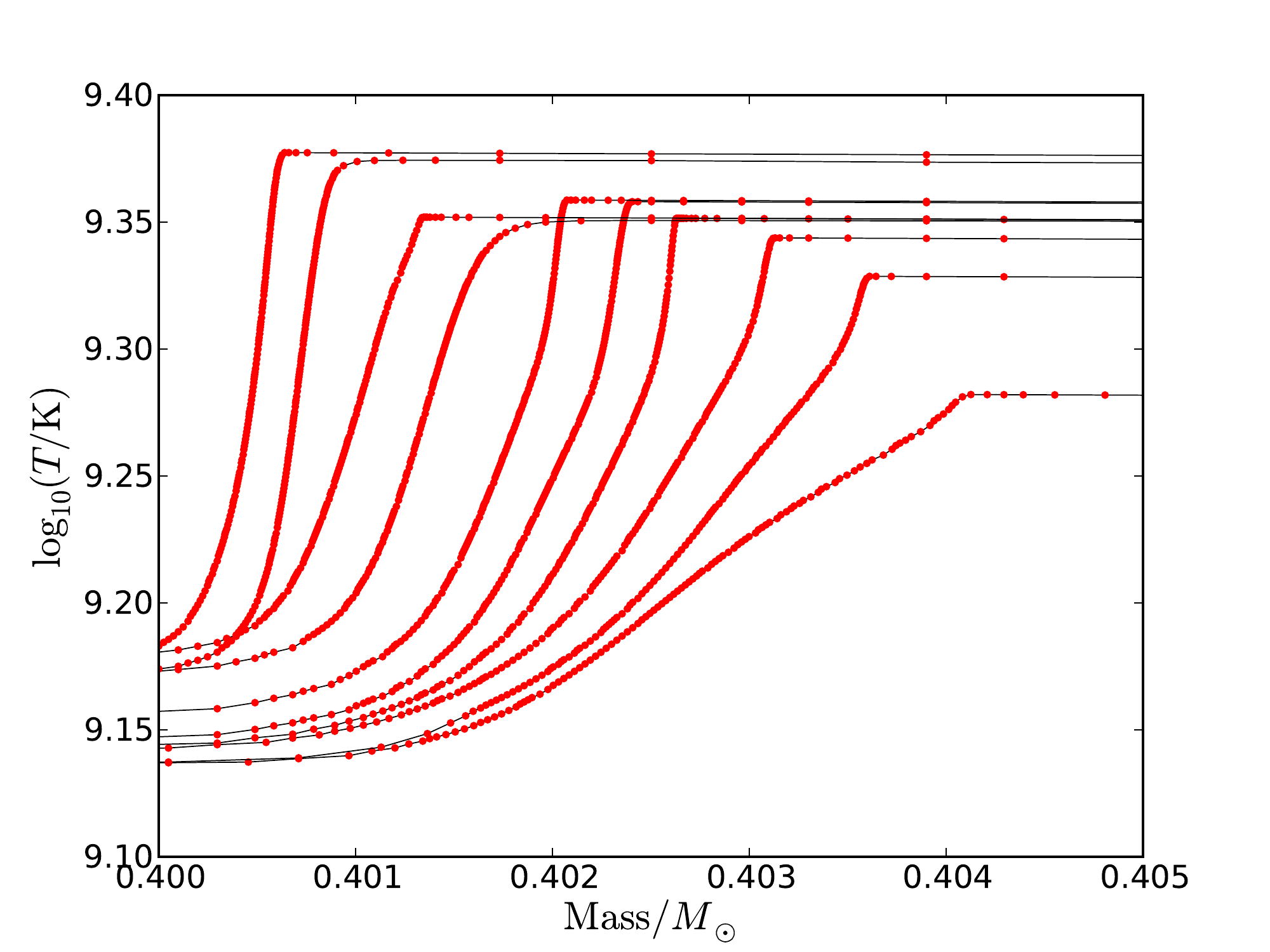}
  \label{}
  }
  \caption{Time evolution of $T$-profiles during the early development of a Ne+O flame. While the nuclear energy produced at the base of the 9.5\,M$_\odot$ flame is deposited into the shells immediately below, allowing for its propagation, the heating effect of the nuclear energy produced in the 8.8\,M$_\odot$ model is diluted due to its transport across a much more distended region by electron conduction owing to higher degeneracy. The flame instead dies away, removing support of the outer layers and allowing the core to contract. Each red dot represents a mesh point in the calculation.}
  \label{flames_Tprofs}
\end{figure*}

Fig.\,\ref{kappa_profiles} shows the opacity profiles following the extinction of
the last neon-oxygen flash and at a later time in each model. Although electron
conduction dominates the heat transfer in both cases, it is more efficient (lower $\kappa$) by a
factor of about 3 in the $8.8M_\odot$ model's early flame and by a factor of
more than 10 later, meaning that any energy production from subsequent radiative
neon-oxygen burning or contraction is diluted across the majority of the core.
In contrast, the higher conductive opacities in the $9.5M_\odot$ model allow for
the nuclear and compressional energy to take effect much more locally, heating
the underlying shell of material to ignition temperatures and causing the
development of a nuclear flame. These two contrasting paths are further illustrated
in Fig.\,\ref{flames_Tprofs}, which shows clearly a flame front developing in the $9.5\,M_\odot$ model
and the dilution of heat across the core of the $8.8\,M_\odot$ model. The effects of spatial resolution on flame development and energy transport are discussed in \S\,\ref{spatial_resolution}.

In summary, the propagation of the flame in the $8.8\,M_\odot$ model is more difficult because of the lower opacity. Furthermore, the combined effect of electron captures in the center and the low $Y_\mathrm{e}$ in the shell due to Ne- and O-burning leads to the core contraction on a shorter timescale than the evolution of the flame.

\subsection{Progenitor structure}
\label{progenitorstructures}
The structure of the progenitor star, in terms of density and electron-fraction profiles of the stellar core, has a strong impact on the timescale at which the later supernova explosion may develop as well as on the explosion energetics. Core-collapse supernova explosions are related to the revival of the stalled shock wave, which forms when the contracting core reaches normal nuclear matter density and bounces back. For massive iron-core stars, the structure of the core at the onset of contraction is determined by the mass enclosed inside the carbon shell. In general, a sharp density gradient separating iron-core and silicon layer results in a strong acceleration of the bounce shock at the onset of shock revival early after core bounce on a timescale of only few 100~ms. Progenitors with a shallower density gradient suffer from a more extended mass accretion period after core bounce, during which the standing bounce shock oscillates, driven by neutrino-energy deposition behind and mass accretion from above. This results in a delayed onset of shock revival by several 100~ms and more energetic explosions due to the larger heat deposition behind the shock via neutrinos before shock expansion. For a recent review of the connection between progenitor structure and recent  axially-symmetric supernova explosion models, see \citet{Janka2012}.

In addition to the standard iron-core progenitors commonly explored in core-collapse supernova studies, we provide a selection of new models of lower zero-age main-sequence mass that belong to the SAGB class as well as to low-mass massive stars. Therefore, in Fig.\,\ref{rhoprof_ne_ignition}, we compare the structures of our SAGB model (8.75~M$_\odot$) after central $^{24}$Mg depletion, electron-capture SN progenitor (8.8~M$_\odot$, failed massive star) at ignition of oxygen deflagration, low-mass massive star (9.5~M$_\odot$) at the point of neon-shell ignition, and standard iron-core progenitor (12~M$_\odot$) at the onset of core contraction. Note that the 9.5~M$_\odot$ progenitor is not then as evolved as the other models and hence its central density is still lower than those of the other models. It is therefore only used as a reference case. The major difference between the low-mass (8.75 and 8.8~M$_\odot$) and the more massive iron-core progenitors is the very steep density gradient separating the core and the envelope. There the density drops about 16 orders of magnitude, from about $10^{8}$ to $10^{-8}$~g~cm$^{-3}$.

Distinguishing the $8.75\,M_\odot$ and $8.8\,M_\odot$ progenitor structures becomes clearer when inspecting the density profiles with respect to radius, Fig.\,\ref{rhoprofradius}. The bulge from $\mathrm{log}_{10}(R/\mathrm{km})\approx3.2$ to $3.8$ that features in the $8.8\,M_\odot$ structure but is absent in the $8.75\,M_\odot$ structure, is a carbon-burning shell. One would expect that, since the $8.8\,M_\odot$ model experienced several neon-oxygen flashes, the structure within the core should be significantly different from that of the super-AGB model. Aside from the abundance profiles showing a large region in which the composition is dominated by Si-group isotopes, the most striking difference is in the electron fraction, $Y_\mathrm{e}$, which is shown in Fig.\,\ref{yeprofs}.

In Fig.\,\ref{rhoprof_ne_ignition} we have included the progenitor structures of the \citet[][SAGB-like]{Nomoto1987} $8.8\,M_\odot$ and the \citet{Woosley2002zz} $12\,M_\odot$ models for comparison. The \citet{Nomoto1987} structure is at a later evolutionary stage compared to our models. A fraction of the core has already been burnt to NSE composition, but the core structure is qualitatively similar to our $8.75\,M_\odot$ SAGB model. It is also clear from Fig.\,\ref{rhoprof_ne_ignition}, right panel, that there are differences in the structure of the \citet[][SAGB-like]{Nomoto1987} model and our $8.8\,M_\odot$ (failed massive star) model, where there is a CO-rich layer at the edge of the core. As discussed previously, there is a neutron-rich layer in our $8.8\,M_\odot$ model where the Ne-O shell flash consumed previously that is not a feature of the \citet{Nomoto1987} model. There is a clear clustering of the SAGB EC-SN progenitor structures and the CCSN progenitor structures in the density profiles as a function of radius (Fig.\,\ref{rhoprof_ne_ignition}, right panel), while the $8.8\,M_\odot$ model lies in-between.

The iron-core progenitors have extended high-density silicon as well oxygen and carbon layers above the core. These result in a shallower transition from iron core to helium envelope. The density decreases steadily step-wise according to the different composition interfaces (see Fig.\,\ref{rhoprof_ne_ignition} left panel). Moreover, different evolutionary tracks for the 8.75, and 8.8~M$_\odot$ progenitor cores lead to low-mass cores of only about 1.376~M$_\odot$, which is significantly lower than for the 12~M$_\odot$ model of $1.89\,M_\odot$ (see Table\,\ref{table_allmodelprops}).
Note that the 12~M$_\odot$ iron-core results are in qualitative agreement with those of the KEPLER code \citep{Woosley2002zz} and, as a function of radius, match very well. The reason for the discrepancy between the two as a function of mass is the difference in assumption for convective overshooting, which has led to the production of larger cores in the MESA model. We are currently working on a code comparison study of MESA, KEPLER and the Geneva stellar evolution code \citep{Hirschi2004} for the evolution, explosion and nucleosynthesis of massive stars in order to quantify some of the related uncertainties.
We expect that 
the resulting steep density gradient at the edge of the core of our EC-SN progenitor models will
accelerate the supernova shock on a short timescale after core bounce, producing a weak explosion with little $^{56}\mathrm{Ni}$ ejecta. Such an explosion should produce qualitatively similar results as obtained for the 8.8~M$_\odot$ progenitor from \citet{Nomoto1987} \citep[for details about electron-capture supernova explosions, see][]{Kitaura2006, Janka2008tobi, Fischer2010}.
The split between weaker, more rapid EC-SN explosions and stronger, slower FeCCSN explosions is a possible explanation for the observed bi-modality in the spin period and orbital eccentricity of X-ray binaries, although it is not clear how this is manifested \citep{Knigge2011}.

\section{Key nuclear and modeling uncertainties}
\label{discussion}
In this section we discuss the main modeling uncertainties affecting the study of stars in the transition mass range. We propose some solutions and suggest ways in which future studies could improve on our models in order to quantify and minimise these uncertainties.
\subsection{Weak reaction rates and the URCA process}
\label{weakdiscussion}
During the very late stages of the $8.75$ and $8.8\,M_\odot$ stars electron captures on
\emph{sd}-shell nuclei become crucial to their fate. In the degenerate
core, there is a very sharp jump in the rates of these electron captures,
which corresponds to a threshold density at which the electron Fermi energy,
$\epsilon_\mathrm{F}$, exceeds the threshold energy for the reaction to proceed.
Tabulated electron capture rates that we use as input for the models must properly
resolve this steep transition if we want to know at what density the oxygen
deflagration is ignited. We should want to know that density so that it can be
determined whether nuclear energy release from burning the core to nuclear
statistical equilibrium (NSE) composition is high enough to exceed the
gravitational binding energy of the core and thus lead to its explosion
\citep{Gutierrez1996}. Otherwise, the core would collapse to a neutron star
following its deleptonisation through electron captures on Fe-group isotopes.

\begin{figure*}
  \centering
  \includegraphics[height=0.44\textheight]{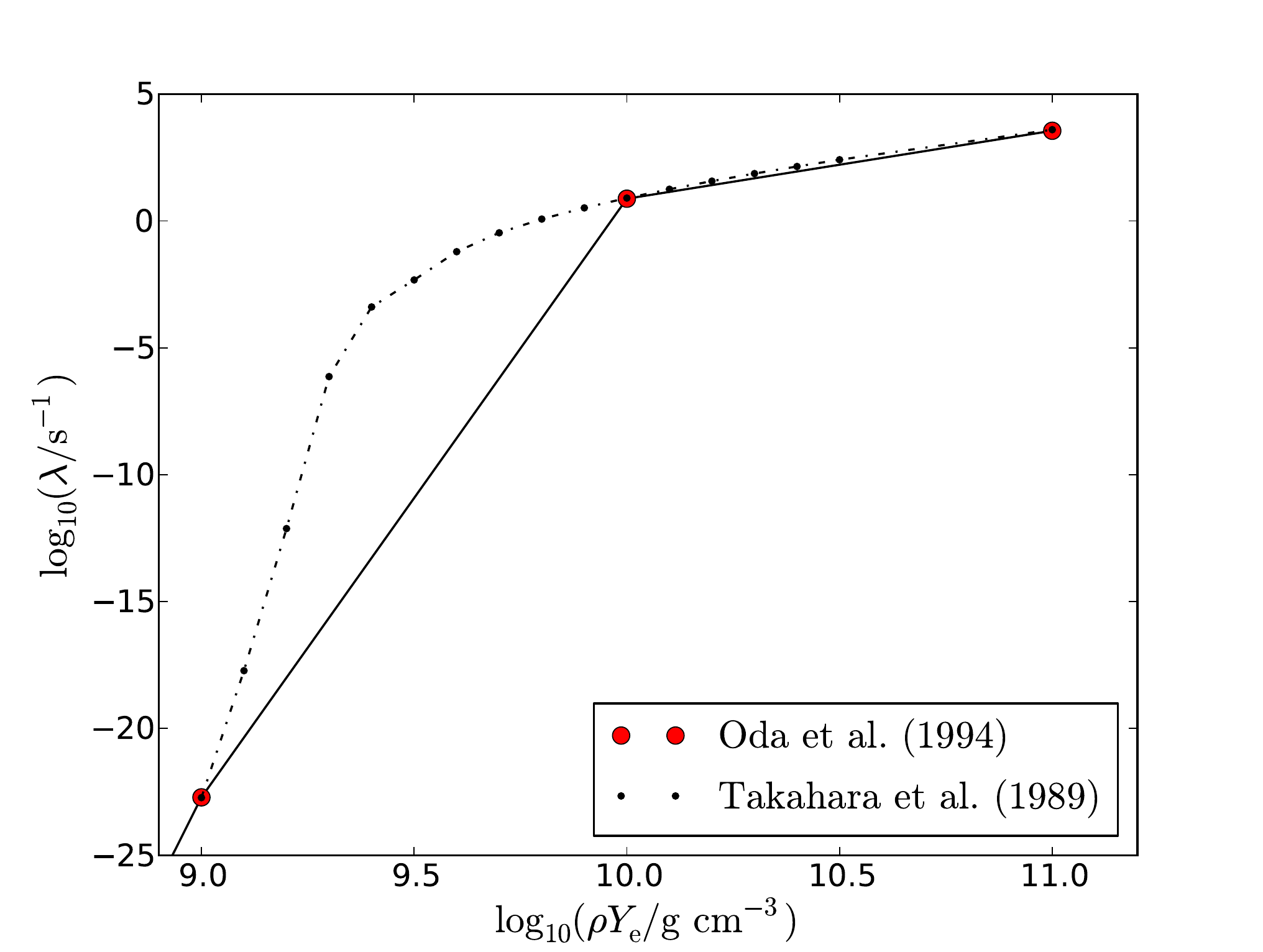}
  \caption{$\lambda(^{24}\mathrm{Mg}+\mathrm{e}^-$) at $T=4\times10^8\,\mathrm{K}$ from the compilations of \citet{ODA94} and \citet{Takahara89}. It is immediately clear that the more favourable sampling in the rate of \citet{Takahara89} better represents the threshold density for the rate. The lines show the resulting interpolation of these rates that is used in the code.}
  \label{oda_vs_takahara_rates}
\end{figure*}

There are still more shortcomings of calculations involving
electron capture rates that are poorly resolved in the $\rho-T$ plane. For example, the vast majority of
widely used rate tables for \emph{sd}-shell nuclei possess a grid spacing of 1
dex in $\rho Y_\mathrm{e}$. As an example one of these crucial reactions,
$^{24}\mathrm{Mg}+\mathrm{e}$, jumps by about 20 orders of magnitude from
$\mathrm{log}_{10}(\rho Y_\mathrm{e}/\mathrm{g\,cm}^{-3})=9.0$ to 10.0 at the temperature of interest
($T\approx0.4~\mathrm{GK}$). This is not only a problem for resolving the rate at
the threshold density, because at lower densities the rate, $\lambda/\mathrm{s}^{-1}$, is significantly
underestimated through linear interpolation of $\mathrm{log}_{10}(\lambda/\mathrm{s}^{-1})$ (see Fig.\,\ref{oda_vs_takahara_rates}).

\begin{figure}
  \centering
  \includegraphics[height=0.44\textheight]{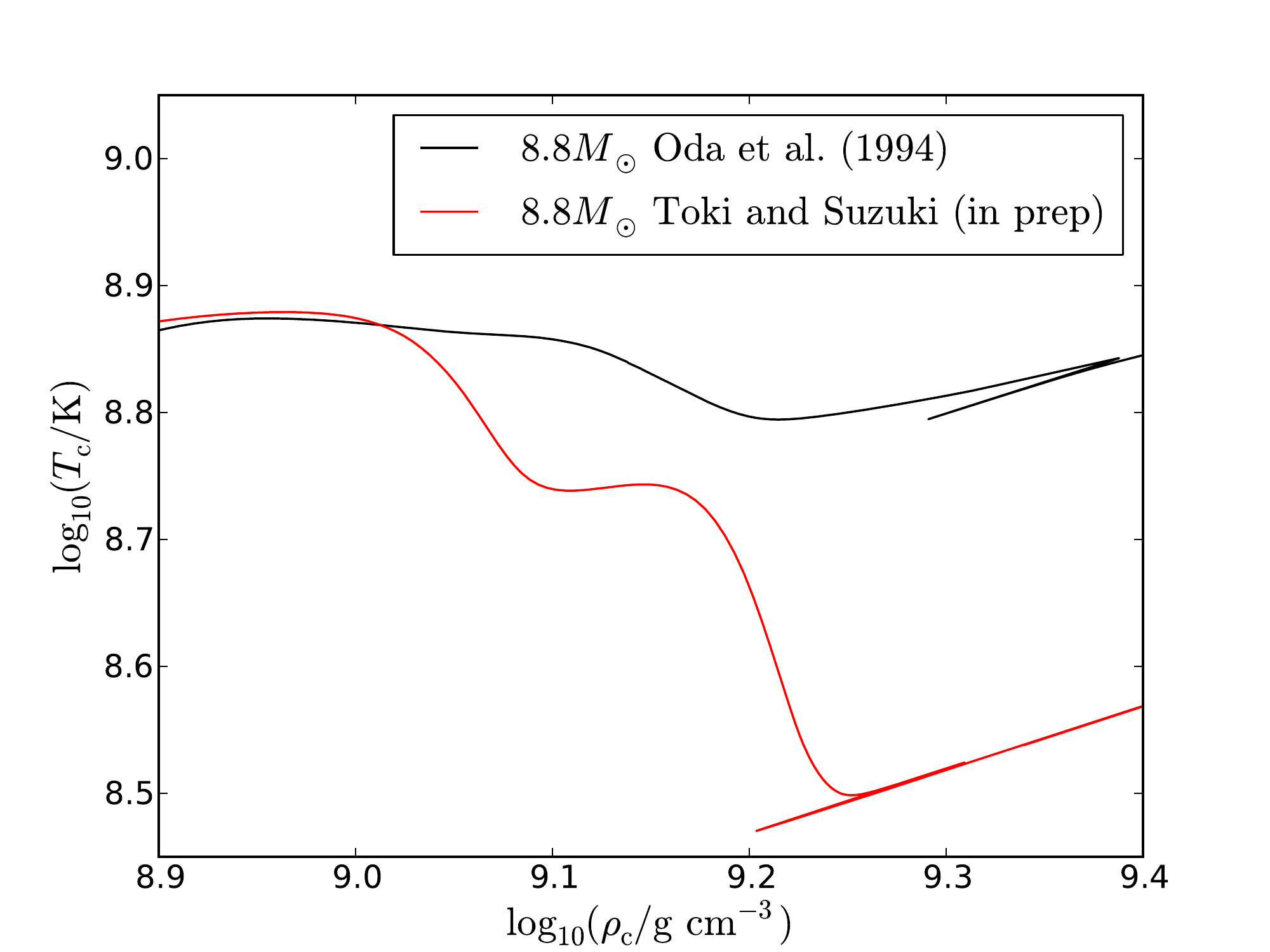}
  \caption{Impacts of the well-resolved rates of Toki \& Suzuki (in prep) on the central evolution of the 8.8\,M$_\odot$ model. The reaction thresholds are more clearly identifiable, and occur at higher densities than with the rates of \citet{ODA94}. The cooling effect is also more pronounced.}
  \label{TokiURCA88tcrhoc}
\end{figure}

It is clear from Fig.\,\ref{tcrhoc_all_models} that the central evolution of the 8.75 and $8.8\,M_\odot$ models is dominated by weak reactions. The onset of the URCA process disrupts the propagation of the neon-oxygen flame and aids the central contraction. In fact, the same is true for all EC-SN progenitors and thus it is imperative to treat the URCA process as accurately as possible to best predict the fate of $8-12\,M_\odot$ stars. \citet{Toki2013} have produced well resolved ($\Delta \mathrm{log}_{10}\,\rho Y_\mathrm{e}/\mathrm{g\,cm}^{-3}=0.02$ and $\Delta \mathrm{log}_{10}\,T/\mathrm{K}=0.05$) reaction and neutrino loss rates for the $A=23, 25$ and $27$ URCA pairs under the conditions $7.0\leq \mathrm{log}_{10}\,(T/\mathrm{K})\leq9.2$ and $8.0\leq \mathrm{log}_{10}\,(\rho Y_\mathrm{e}/\mathrm{g\,cm}^{-3})\leq9.2$. The differences between these new rates and those of \citet{ODA94} is shown in Fig.\,\ref{toki_vs_oda_rates} for the $A=25$ pair at $T_9=0.4$. The impacts of these new, well-resolved rates compared to those of \citet{ODA94} are shown in Fig.\,\ref{TokiURCA88tcrhoc}. Not only is the cooling effect more pronounced, the reaction thresholds are more clearly identifiable and occur at higher densities than with the rates of \citet{ODA94}.

Because the rates are so sensitive to density, any form of interpolation cannot
properly represent the physical situation without some input from knowledge of
the nuclear physics. This is why several groups employ an interpolation of
effective $\mathrm{log}\,ft$ values \citep{FFNweak1985}. An effective $\mathrm{log}\,ft$ value for a reaction is related to its
raw rate by the relationship in Eq.\,\ref{logftrelation},
\begin{equation}
 ft=\frac{\phi}{\lambda}\,\mathrm{ln}\,2
 \label{logftrelation}
\end{equation}
where $\phi$ is the ground-state to ground-state phase space integral. The aim
is to produce a quantity that varies smoothly with $T$ and $\rho$ from which the
raw rate may be obtained within a stellar evolution calculation by approximation
of the phase space integral at the desired conditions. This method is relatively
robust for those weak rates for which ground-state to ground-state transitions
dominate. However this is not the case for the reactions of interest in
electron-capture supernova progenitors. The change in $Y_\mathrm{e}$ is not the
only important facet of the electron captures; they also posses a strong heating
effect due to the $\gamma$-decay following transitions to excited
states of the daughter nuclei. Hence this demonstrates the importance of excited
states when we attempt to normalise the reaction rate using simplifications or approximations.

Therefore we conclude that there are two possible sets of desired quantities, either grids of
weak reaction rates for \emph{sd}-shell nuclei that are appropriately resolved
through the threshold density or $\mathrm{log}ft$ values that incorporate all
important transitions in the normalisation of the rate. There are
contributions from many states of the parent and daughter nuclei for these
reactions and to perform phase space integral routines within a stellar
evolution code to account for this could be exceptionally inefficient. It is also important to use $\beta^\pm$-decay and neutrino-loss rates calculated with the same physics and grid resolution to ensure consistency when we look at the impact of the URCA process on the evolution. The most up-to-date rates should also include the effects of Coulomb screening, which has been shown to increase the threshold density for electron captures \citep{Gutierrez1996}. An increase in the threshold density of $^{20}\mathrm{Ne}(\mathrm{e}^-,\nu)^{20}\mathrm{F}$ would cause the oxygen deflagration to ignite under denser conditions in the super-AGB progenitors. However in the failed massive star case the center is approaching the ignition temperatures of Ne and O adiabatically, and so the oxygen deflagration could be ignite before $^{20}\mathrm{Ne}+\mathrm{e}^-$ becomes significant if there were an increase in the threshold density.

\subsection{Uncertainties due to convection}
\begin{figure}
  \centering
  \includegraphics[height=0.44\textheight]{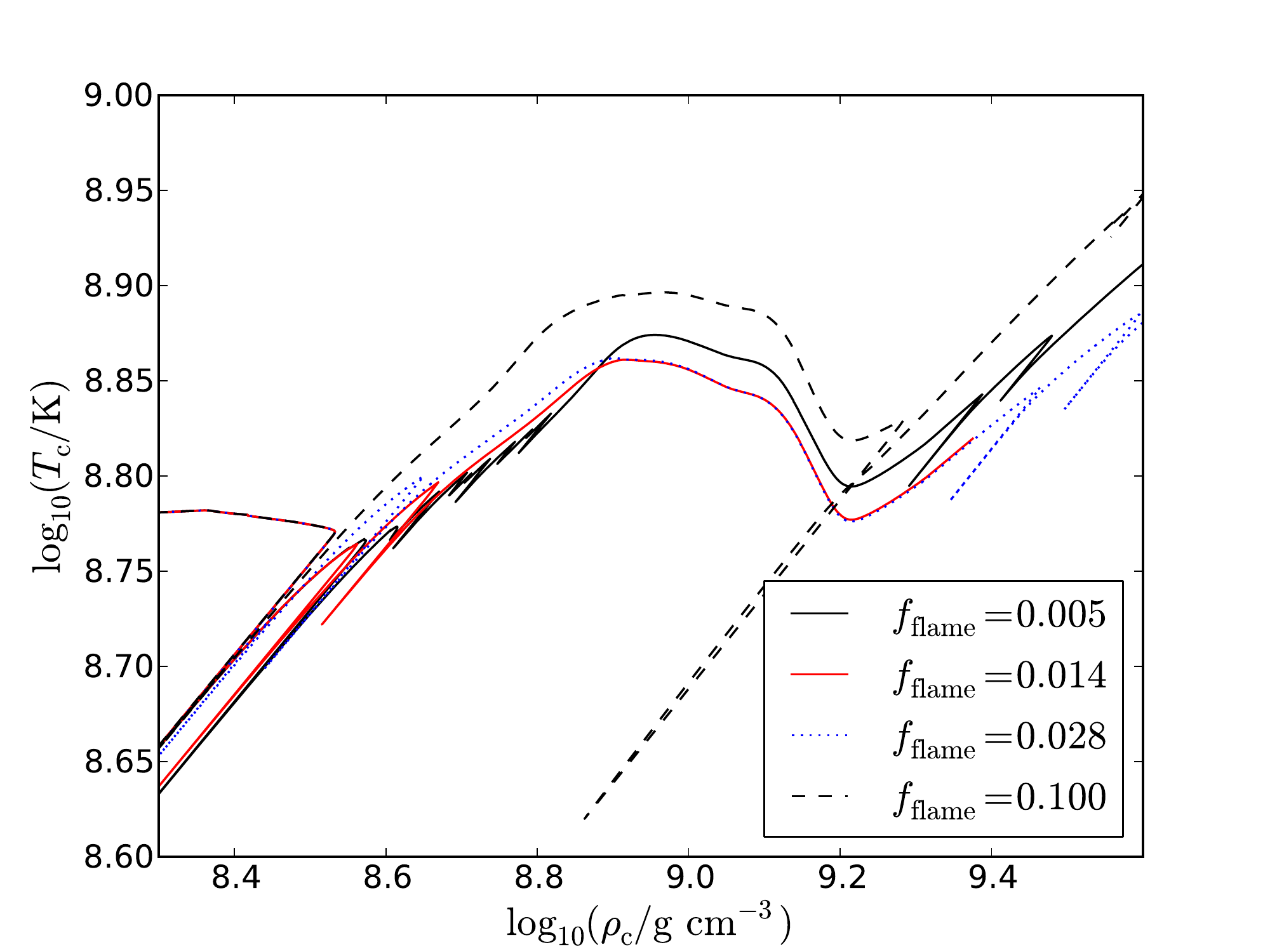}
  \caption{Central density--temperature evolution of the 8.8\,M$_\odot$ model showing the differences created when we assumed $f_\mathrm{flame}=0.005$ (our original
assumption), 0.014, 0.028 and 0.100 (extreme), where $f_\mathrm{flame}$ is the value of
the parameter $f$ in Eq.\,\ref{exponentialovershoot} at the base of the
ONe-burning shell.}
  \label{88_fzbelow_tcrhoc_fig}
\end{figure}

Still one of the largest uncertainties in any 1-D stellar evolution calculation is the
treatment of convection. Extra mixing at convective boundaries may explain many observed phenomena, for example the s-process abundance patterns
in AGB stars, and hence we include such mixing in our models. Due to the
turbulent and advective nature of convection, it is physically plausible to infer
some extra mixing across the boundary between convective and radiative layers but without the benefit of 3-D hydrodynamical
simulations of the physical conditions it is difficult to quantify its extent.
We use the term convective boundary mixing rather than
overshooting for the advanced evolution phases of the deep stellar interior, such as convective shells. This is because the term overshooting suggests a physical picture in which coherent convective structures or blobs cross the Schwarzschild boundary before they notice the reversal of buoyancy acceleration. However, in the deep interior other hydrodynamic instabilities, such as Kelvin-Helmholtz or internal gravity wave induced turbulence dominate mixing at the convective boundary.
Largely, the effect of including convective boundary mixing is to shift the
transition masses due to increased core sizes. However it is intuitive to
hypothesise that increased amounts of extra mixing below the ONe-burning shells
would have a crucial effect on their inward propagation. To test this, we assumed extra
mixing below the convective ONe-burning shells to behave as an
exponentially decaying diffusion process as outlined in
Eq.\,\ref{exponentialovershoot} with $f_\mathrm{flame}=0.005$ (our original
assumption), 0.014, 0.028 and 0.100. The central density--temperature evolution
from the flame's ignition for all of these assumptions is shown in
Fig.\,\ref{88_fzbelow_tcrhoc_fig}. It should be noted that setting $f_\mathrm{flame}=0.100$ is an extremely unphysical assumption that we adopt simply to test the uncertainty of our conclusions.

\begin{figure}
  \centering
  \subfigure[$f_\mathrm{flame}=0.005$]{
  \includegraphics[height=0.44\textheight]{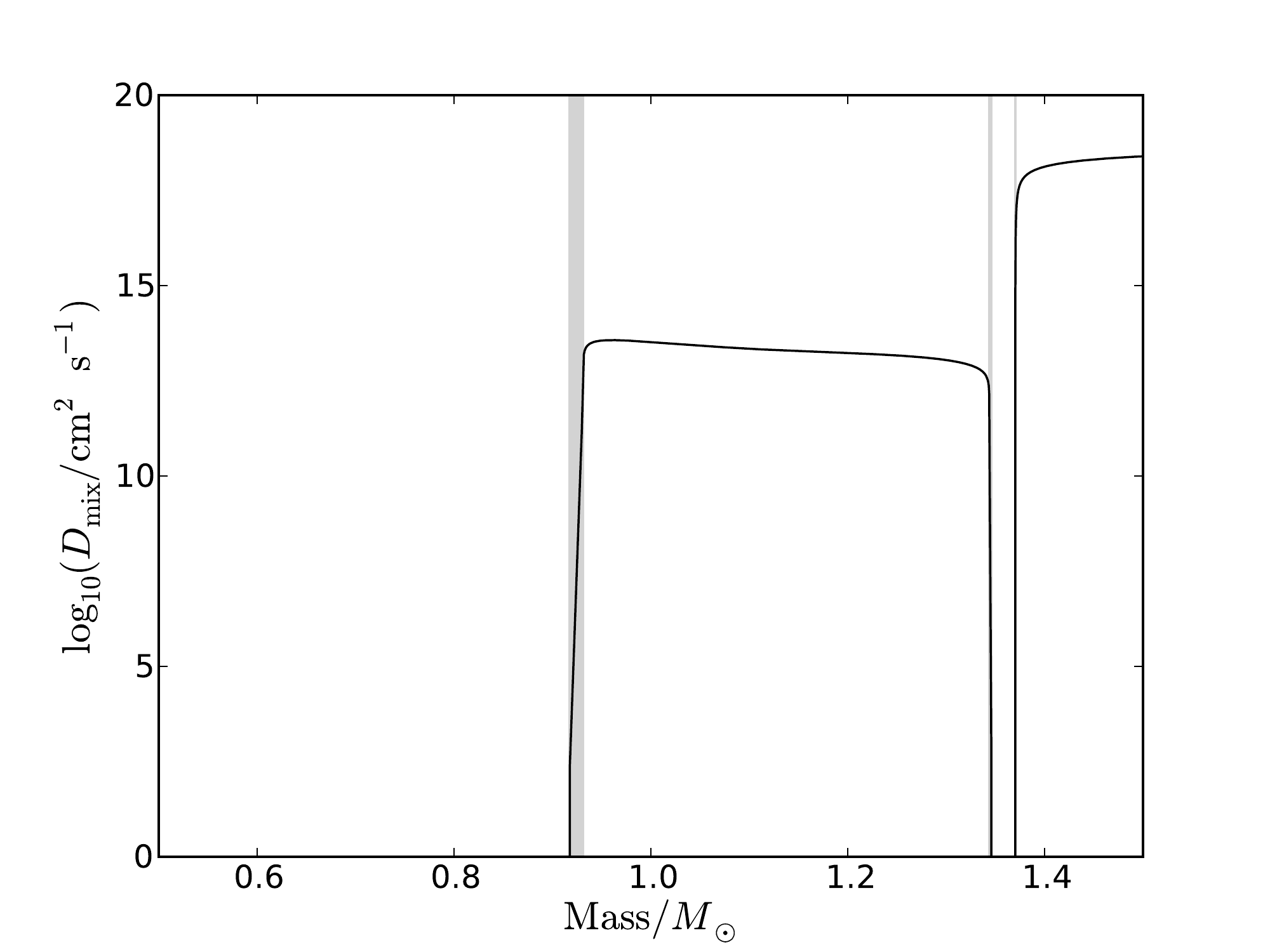}
  \label{88_fz005_dmixCBM}
  }
  \subfigure[$f_\mathrm{flame}=0.100$]{
  \includegraphics[height=0.44\textheight]{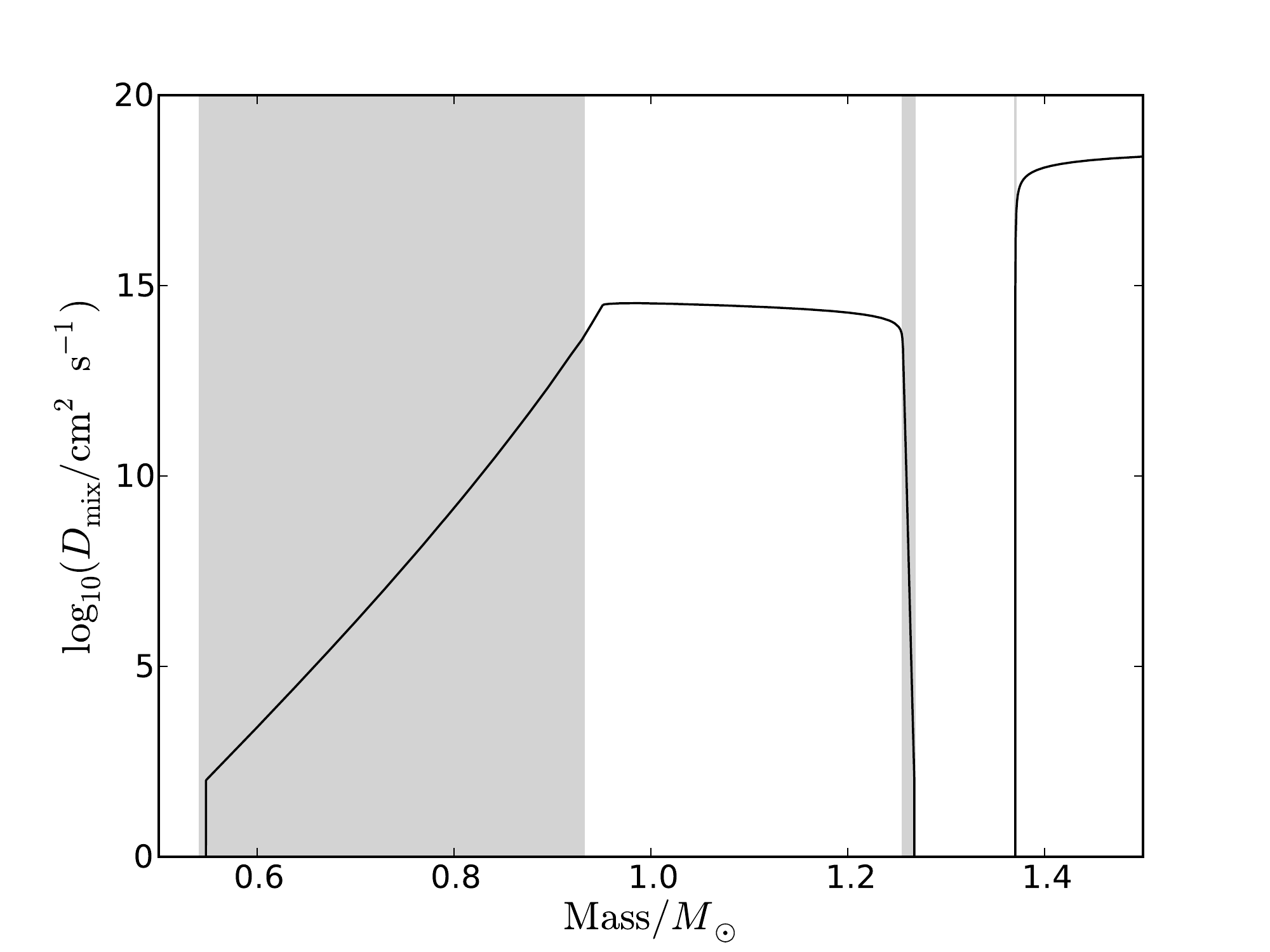}
  \label{88_fz100_dmix_CBM}
  }
  \caption{Radial diffusion coefficient profiles against mass coordinate during
the second neon flash event in the 8.8\,M$_\odot$ model. Shaded grey areas represent the regions of convective boundary mixing.
Although the flame re-ignites in the case with $f_\mathrm{flame}=0.100$, the
fuel is brought in on the mixing timescale which, during this phase, is shorter than
the central contraction timescale and the critical density is already reached
for $^{24}\mathrm{Mg}+\mathrm{e}^-$, leaving the outcome of the model
unaltered. It should also be noted that $f_\mathrm{flame}=0.100$ is an extreme assumption adopted purely for the purpose of testing the robustness of our models.}
  \label{dmix_plots}
\end{figure}

Although the central evolution behaves slightly differently for each mixing
assumption, all the models reach central densities of $\rho_\mathrm{c} =
10^{9.6}$ g cm$^{-3}$ at temperatures well below the neon-ignition threshold. The model with
the largest amount of mixing ($f_\mathrm{flame}=0.100$) is unique because
although all models undergo a few flashes after the extent of the URCA process
has been exhausted, it is the only one to re-ignite an ONe shell at a mass
co-ordinate in-keeping with the original location of the flashes\footnote{All of
the other models in this test ignite further shells at the locus of maximum
extent of the original ONe-shell flashes, similar to the standard $8.8M_\odot$
case.}. At this point, the center is already extremely close to the threshold
density for $^{24}\mathrm{Mg}(\mathrm{e}^-,\nu)^{24}\mathrm{Na}$ at
$\rho_\mathrm{c}\approx10^{9.6}$ g cm$^{-3}$. The change in extent of the convective boundary
mixing between the $f_\mathrm{flame}=0.005$ and $f_\mathrm{flame}=0.100$ models
is shown in Fig.\,\ref{dmix_plots}. The deep mixing in the extreme
($f_\mathrm{flame}=0.100$) model replenishes fuel at the flame front, allowing it
to re-ignite.

Convective boundary mixing is at present still a very uncertain phenomenon. While the timescales for stellar evolution restrict theoretical models to only one dimension, there is an emergence of effort to explore specific phases of the evolution in two \citep{Herwig2006,Herwig2007} and three \citep{Mocak2011,Herwig2011} dimensions in order to properly quantify the extent of convective mixing and its behaviour at the boundary with a radiative zone. We plan to perform an in-depth parameter study of convective boundary mixing in $8-10\,M_\odot$ stars. The long term goal is to constrain the parameters of our diffusive treatment by analysing the data from 3D simulations.

\subsection{Spatial resolution of the ONe flame}
\label{spatial_resolution}
The development and propagation of a nuclear flame front is highly sensitive to
the spatial resolution due to the thin flame width. Fig.\,\ref{flames_Tprofs} showed the evolution of the $T$-profile at the time of flame development/propagation in both models. Each red dot represents a mesh point in the calculation. It can be seen that the model possesses a spatial resolution much finer than the width of the flame front, however in the transition at the base of the flame it is evident that there is a less than desirable resolution very early on in its development.

To examine the effect of spatial resolution on the outcome of the $8.8M_\odot$ model, we increased the resolution of the model tenfold and then twenty-fold at the base of the Ne+O-burning convection zone, from before the ignition of the Ne+O flame. In a second test we increased the resolution tenfold in the regions where energy production from $^{16}\mathrm{O}(\alpha,\gamma)^{20}\mathrm{Ne}$, $^{20}\mathrm{Ne}(\alpha,\gamma)^{24}\mathrm{Mg}$, $^{16}\mathrm{O}(^{16}\mathrm{O},\gamma)^{32}\mathrm{S}$, $^{16}\mathrm{O}(^{16}\mathrm{O},p)^{31}\mathrm{P}$ or $^{16}\mathrm{O}(^{16}\mathrm{O},\alpha)^{28}\mathrm{Si}$ became significant (greater than $10^{4}~\mathrm{erg\,g}^{-1}\mathrm{ s}^{-1}$).
Neither of the enhanced resolutions at the base of the convective shell alter the outcome of the $8.8\,M_\odot$ model (EC-SN), and nor does the re-meshing based on energy production. This demonstrates that our results concerning Ne-O flame propagation are robust and not due to an under-resolved flame front.

\section{Discussion and concluding remarks}
\label{conclusions}

We have begun to explore in detail the transition mass between super-AGB
stars and massive stars. Using the MESA code, we were able to model
stars across the transition (AGB, SAGB, EC-SNe progenitors and massive stars)
with a consistent set of input physics, while current published stellar evolution calculations
limit themselves to either massive stars or SAGB stars.

We were able to follow the evolution of the entire star from pre-MS up to
the ignition of an oxygen deflagration for the $8.8\,M_\odot$ model and up to electron captures on $^{20}\mathrm{Ne}$ for the $8.75\,M_\odot$ model, both of which become EC-SNe. The $8.75\,M_\odot$ case is the first EC-SN progenitor model published including the envelope and the TP-SAGB phase, and the $8.8\,M_\odot$ case is the first EC-SN progenitor model from a failed massive star. Using the AGILE-BOLTZTRAN
hydrodynamics code, we confirmed the $8.8\,M_\odot$ model to result in core collapse --- an EC-SN \citep{FischerECSN2013}.
Our models confirm the notion that failure to establish a stable neon-oxygen laminar flame that propagates to the center can result in an electron-capture supernova. The main difference in the pre-SN evolution when compared with
the generally accepted (SAGB) EC-SN progenitors is that following dredge-out (and neon-burning),
the core contracts directly to the 
threshold density for electron captures by $^{24}\mathrm{Mg}$ and $^{20}\mathrm{Ne}$
as opposed to first undergoing episodic core growth through thermal pulses in the He-shell.

Although the main conclusions of this paper should not change, we need to stress that the initial mass for which the evolutionary paths described in this letter take place depend on the choices made for convective boundary mixing. We also note that the rates of weak reactions in \emph{sd}-shell nuclei of \citet{ODA94} are available in very sparse grids with respect to temperature and electron density. Finer grids for weak interaction rates are necessary to precisely follow the URCA
and other weak reaction processes. We have shown that by using new, well sampled weak rates for the $A=23$, 25 and 27 URCA pairs we reproduce qualitatively the central evolution presented by \citet{Ritossa1999}.

\begin{figure*}
  \centering
  \subfigure[]{
  \includegraphics[width=0.7\textwidth]{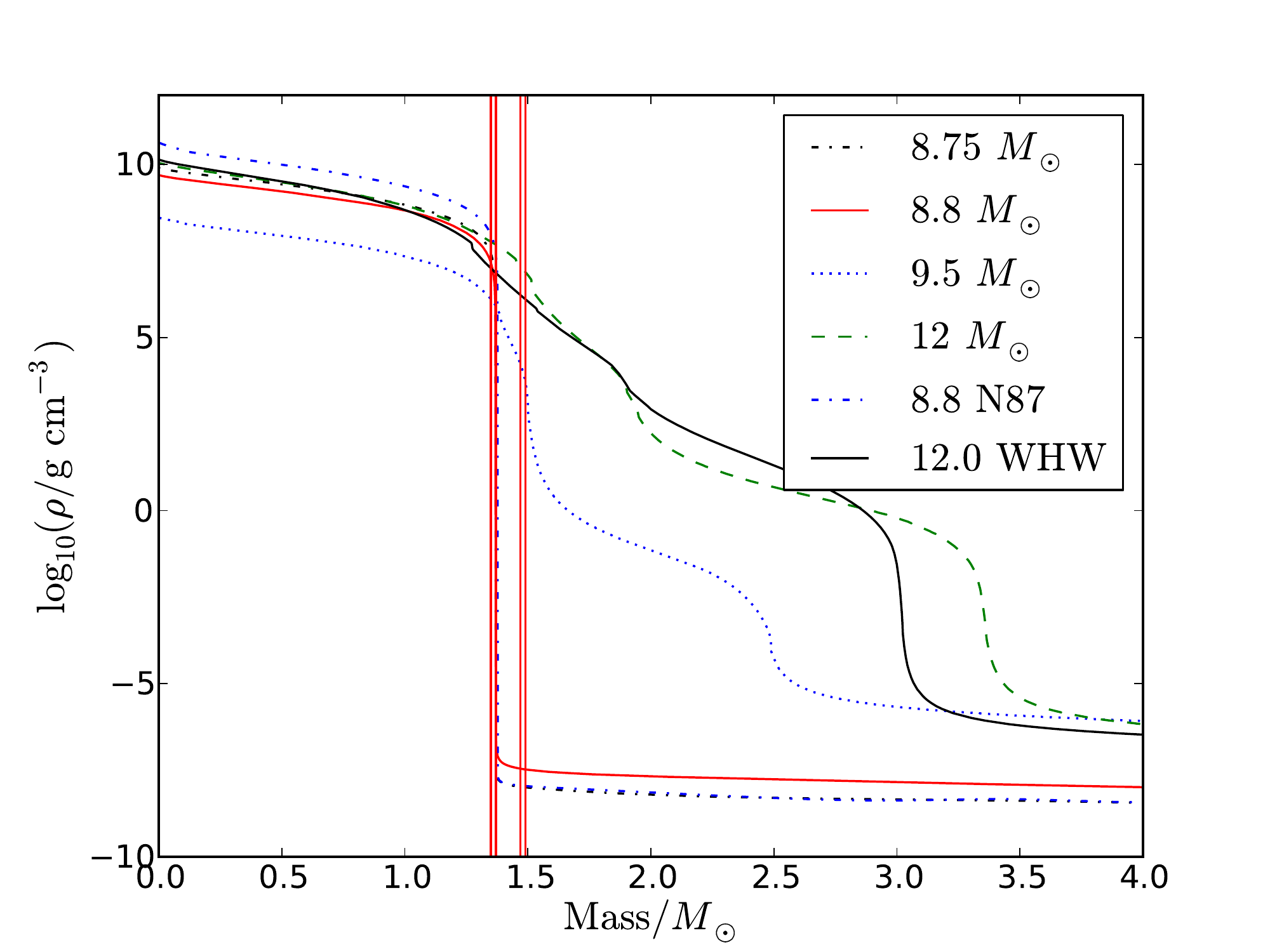}
  \label{rhoprofmass}
  }
  \subfigure[]{
  \includegraphics[width=0.7\textwidth]{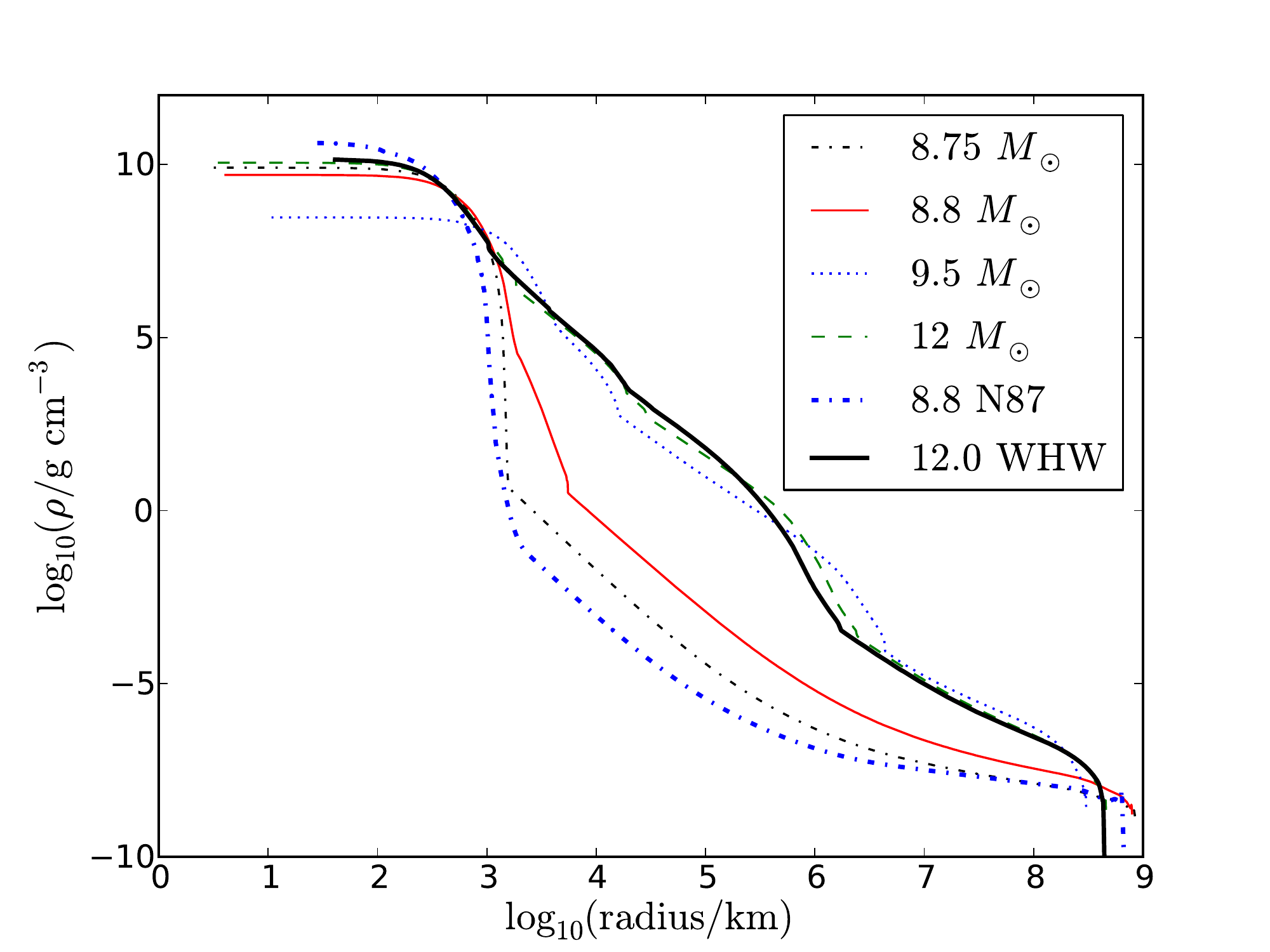}
  \label{rhoprofradius}
  }
  \caption{Density profiles as a function of mass coordinate (a) and radius (b) for 8.75, 8.8 and 12\,M$_\odot$ models after central $^{24}\mathrm{Mg}$ depletion, ignition of oxygen deflagration and collapse, respectively. The 9.5\,M$_\odot$ density profile at the point of neon ignition is also plotted for reference. While the 8.8\,M$_\odot$ model possesses an SAGB-like
  structure following dredge-out, the 9.5\,M$_\odot$ is more reminiscent of a
  massive star with distinct He- and C- shells. Vertical red lines in (a) show
  derived pre-collapse masses for the
  two peaks in the observed neutron star distribution of \citet{Schwab2010}. The blue dot-dashed line shows the structure of the \citet{Nomoto1987} progenitor and the black solid line shows that of the 12\,M$_\odot$ progenitor from \citet{Woosley2002zz}.}
  \label{rhoprof_ne_ignition}
\end{figure*}

\citet{Schwab2010} present a sample of 14 neutron stars for which the masses are well-measured. The authors
calculate the pre-collapse masses of the stars in their sample, the distribution of which is distinctly bimodal and
is attributed to the two birth mechanisms, EC-SNe and FeCCSNe. In Fig.\,\ref{rhoprof_ne_ignition}, the two
peaks of the pre-collapse mass distribution are plotted as red vertical lines.
Because each NS birth mechanism is coupled intrinsically to the pre-SN evolution of the star, it is an
interesting result that the peaks should agree relatively well with the pre-SN structure of the two models in
our set that undergo off-center ignition of neon.
Between $8.8\,M_\odot$ and $9.5\,M_\odot$ (from our two models), an initial mass range of only $0.7\,M_\odot$ contains about $15\%$ of
all single stars with the potential to give birth to a NS (assuming a Salpeter IMF and that single stars in the mass range
$8.5\leq M/M_\odot\leq20$ produce neutron stars in their deaths). For this reason, we stress the importance
of further investigation into the initial mass range between $8.8\,M_\odot$ and $9.5\,M_\odot$. From examination of these two models in our set, there may be an interesting correlation between the propagation of the neon-oxygen flame and the URCA process.

If both failed massive stars and super-AGB stars have the potential to produce electron-capture supernovae
then the EC-SN channel is wider than we think at present. It is our intention to produce EC-SN progenitor models from both super-AGB stars and failed massive stars for several metallicities. Detailed supernova simulations with our models and including
full nucleosynthesis will help constrain what observational features and nucleosynthesis we can expect from EC-SNe.

\acknowledgments

The research leading to these results has received funding from the European Research Council under the European Union's Seventh Framework Programme (FP/2007-2013)/ERC Grant Agreement n. 306901.
NuGrid acknowledges significant support from NSF grants PHY 02-16783 and PHY 09-22648 (Joint Institute for Nuclear Astrophysics, JINA). 
R. H. thanks the Eurocore project Eurogenesis for support.
K. N., R. H. and S. J. acknowledge support from the World Premier International Research Center Initiative (WPI Initiative), MEXT, Japan.
T. F. acknowledges support from the Swiss National Science Foundation under project no. PBBSP2-133378 and HIC for FAIR.
B. P.'s  research has been supported by the National
Science Foundation under grants PHY 11-25915 and AST 11-09174.
M. G. B.'s research was carried out under the auspices of the National Nuclear Security Administration of the U.S. Department of Energy at Los Alamos National Laboratory under Contract No. DE-AC52-06NA25396.

\bibliographystyle{apj}

\clearpage

\begin{deluxetable*}{rcccccc}
\tablecolumns{7}
\tabletypesize{\scriptsize}
\setlength{\tabcolsep}{0.02in}
\tablewidth{0.7\textwidth}
\tablecaption{Summary of model properties}
\tablehead{
\colhead{} & \colhead{8.2\,M$_\odot$} & \colhead{8.7\,M$_\odot$} & \colhead{8.75\,M$_\odot$} & \colhead{8.8\,M$_\odot$} & \colhead{9.5\,M$_\odot$} & \colhead{12.0\,M$_\odot$}
}
\startdata
$M^\mathrm{C}_\mathrm{ign}/M_\odot\,^a$   & 0.15 & 0.00 & 0.00 & 0.00   & 0.00   & 0.00   \\
$M^\mathrm{Ne}_\mathrm{ign}/M_\odot\,^b$  & --   & -- & -- & 0.93 & 0.42 & 0.00 \\
$T^\mathrm{Ne}_\mathrm{ign}/GK\,^c$  & -- & -- & -- & 1.318 & 1.311 & 1.324 \\
$\psi^\mathrm{Ne}_\mathrm{c}\,^d$ & -- & -- & -- & 46.0 & 15.2 & 5.6 \\
$\rho^\mathrm{Ne}_\mathrm{c}/\mathrm{g cm}^{-3}\,^e$ & -- & -- & -- & $3.343\times10^8$ & $7.396\times10^7$ & $1.730\times10^7$\\
$M_\mathrm{tot}/M_\odot\,^f$ & 7.299   & 7.910   & 8.572 & 8.544   & 9.189   & 11.338    \\
$M_\mathrm{env}/M_\odot\,^g$ & 6.031   & 6.559   & 7.210 & 7.174   & 6.702   & 8.023     \\
$M_\mathrm{He}/M_\odot\,^h$  & 1.26721 & 1.35092 & 1.36230 & 1.36967 & 2.48733 & 3.31580   \\
$M_\mathrm{CO}/M_\odot\,^i$  & 1.26695 & 1.35086 & 1.36227 & 1.36964 & 1.49246 & 1.88602   \\
\\
Remnant & ONe WD & ONe WD / NS & -- & NS & NS & NS  \\
SN Type & -- & -- / EC-SN (IIP) & -- & EC-SN (IIP) & CC-SN (IIP) & CC-SN(IIP)  \\
\enddata
\label{table_allmodelprops}
\tablecomments{$^a$ Mass coordinate of carbon ignition. $^b$ Mass coordinate of neon ignition. \\
$^c$ Temperature at locus of neon ignition. $^d$ Central degeneracy at time of neon ignition. \\
$^e$ Central density at time of neon ignition. \\
$^f$ Total mass at time of first thermal pulse or neon ignition. \\
$^g$ Envelope mass at time of first thermal pulse or neon ignition. \\
$^h$ Helium core mass (H-free core mass) at time of first thermal pulse or neon ignition. \\
$^i$ Carbon-oxygen core mass (He-free core mass) at time of first thermal pulse or neon ignition.
}
\end{deluxetable*}

\begin{figure*}
  \centering
  \includegraphics[height=0.44\textheight]{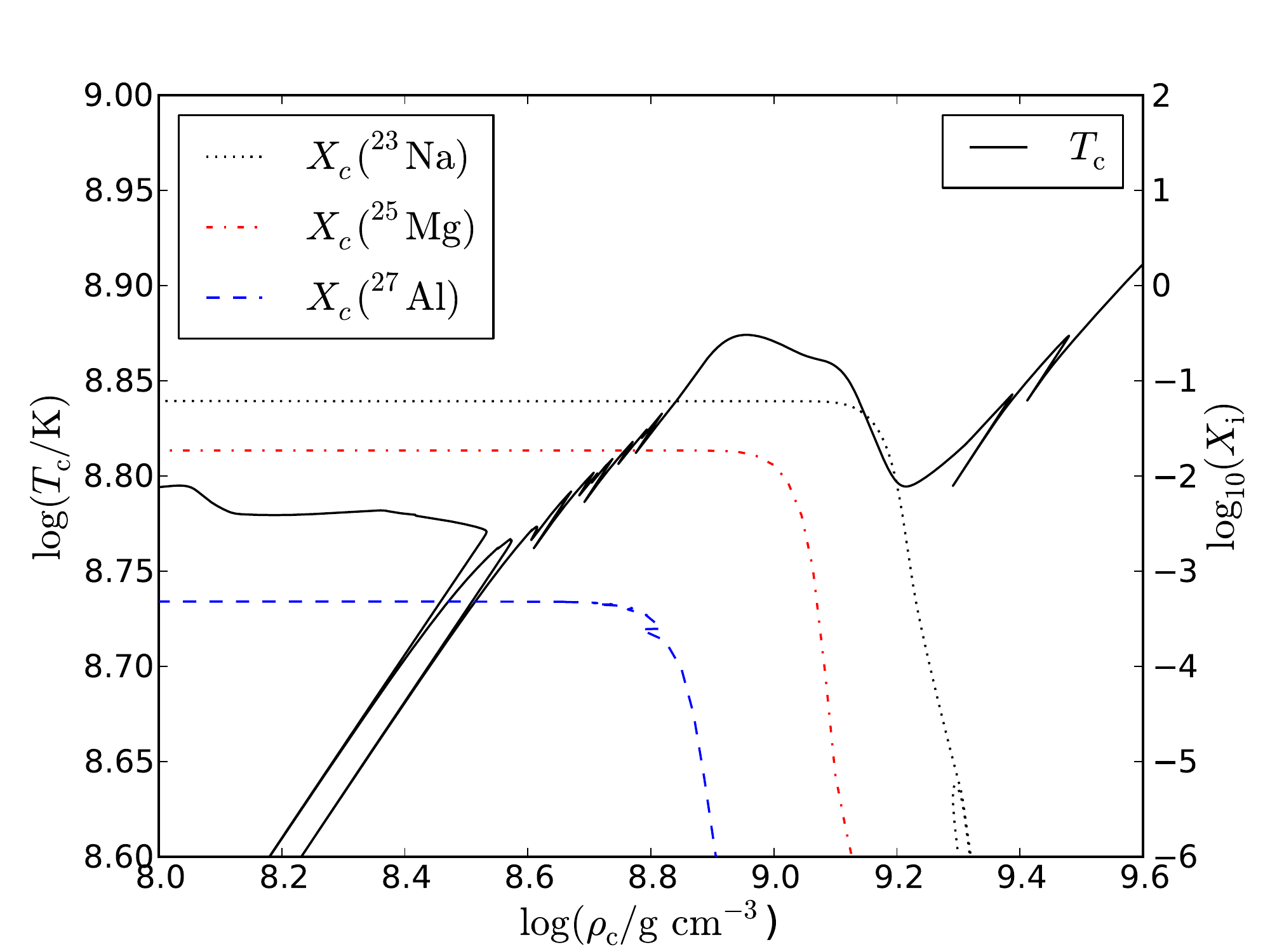}
  \caption{Evolution of the 8.8\,M$_\odot$ model in the
$\rho_\mathrm{c}-T_\mathrm{c}$ plane along with the central abundances (right axis) of the
key URCA process isotopes. Electron captures on $^{25}\mathrm{Mg}$ and
$^{23}\mathrm{Na}$ cool the central regions while those on $^{27}\mathrm{Al}$
provide little contribution due to the low abundance of fuel and the Pauli
blocking of $^{27}\mathrm{Mg}\rightarrow^{27}\mathrm{Al} + \beta^- + \bar{\nu}$.}
  \label{88_URCA_tcrhoc}
\end{figure*}

\end{document}